\PassOptionsToPackage{utf8}{inputenc}
\documentclass[preprint, 3p]{elsarticle}
\usepackage{microtype}
\usepackage{graphicx,subfigure}
\usepackage{booktabs} % for professional tables
\usepackage{diagbox}
\usepackage{dsfont}
\usepackage{url}
\usepackage[ruled,vlined, noend]{algorithm2e}
\usepackage{amsmath, amsthm, amssymb, amsfonts}
\usepackage{array}
\usepackage{shortbold}
\usepackage{xcolor}
\usepackage[title]{appendix}
\usepackage[capitalise]{cleveref}
\usepackage{xr}
\usepackage{adjustbox}
\usepackage{ulem}
\usepackage{cancel}
\usepackage{natbib}
\renewcommand{\appendixname}{Supplementary}

\setcitestyle{authoryear, round}
\begin{document}
\begin{frontmatter}
\title{\texttt{AntBO}: Towards Real-World Automated Antibody Design with Combinatorial Bayesian Optimisation}
\author[2]{Asif Khan\corref{cor1}\fnref{fn1}}
\address[1]{Huawei Noah{'}s Ark Lab, London, N1C 4AG, United Kingdom}
\address[2]{School of Informatics, University of Edinburgh, Edinburgh, EH8 9YL, United Kingdom}
\author[3]{Alexander I. Cowen-Rivers\corref{cor1}\fnref{fn1}}
\address[3]{Intelligent Autonomous Systems, Technische Universität Darmstadt, Darmstadt, 64289, Germany}
\author[1]{Antoine Grosnit}
\author[1]{Derrick-Goh-Xin Deik}
\author[4]{Philippe A. Robert}
\address[4]{Department of Immunology, University of Oslo, Oslo, 0315, Norway}
\author[4]{Victor Greiff}
\author[4]{Eva Smorodina}
\author[4]{Puneet Rawat}
\author[1]{Kamil Dreczkowski}
\author[4]{Rahmad Akbar}
\author[1]{Rasul Tutunov}
\author[5]{Dany Bou-Ammar}
\address[5]{American University of Beirut Medical Centre, Beirut, 11-0236, Lebanon}
\author[1,6]{Jun Wang}
\address[6]{University College London, London, WC1E 6BT, United Kingdom}
\author[2]{Amos Storkey}
\author[1,6]{Haitham Bou-Ammar}
\cortext[fn1]{These authors Contributed Equally}
\fntext[cor1]{Corresponding Author}
\begin{abstract}
Antibodies are canonically Y-shaped multimeric proteins capable of highly specific molecular recognition. The CDRH3 region located at the tip of variable chains of an antibody dominates antigen-binding specificity. Therefore, it is a priority to design optimal antigen-specific CDRH3 regions to develop therapeutic antibodies. However, the combinatorial nature of CDRH3 sequence space makes it impossible to search for an optimal binding sequence exhaustively and efficiently using computational approaches. Here, we present \texttt{AntBO}: a combinatorial Bayesian optimisation framework enabling efficient \textit{in silico} design of the CDRH3 region. Ideally, antibodies are expected to have high target specificity and developability. We introduce a CDRH3 trust region that restricts the search to sequences with favourable developability scores to achieve this goal. For benchmarking, \texttt{AntBO} uses the \texttt{Absolut!} software suite as a black-box oracle to score the target specificity and affinity of designed antibodies \textit{in silico} in an unconstrained fashion~\citep{robert2021one}. The experiments performed for $159$ discretised antigens used in \texttt{Absolut!} demonstrate the benefit of \texttt{AntBO} in designing CDRH3 regions with diverse biophysical properties. In under $200$ calls to black-box oracle, \texttt{AntBO} can suggest antibody sequences that outperform the best binding sequence drawn from 6.9 million experimentally obtained CDRH3s and a commonly used genetic algorithm baseline. Additionally, \texttt{AntBO} finds very-high affinity CDRH3 sequences in only 38 protein designs whilst requiring no domain knowledge. We conclude \texttt{AntBO} brings automated antibody design methods closer to what is practically viable for in vitro experimentation.
\end{abstract}
% \begin{graphicalabstract}
% %\includegraphics[width=0.98\textwidth, trim={0 0 6cm 0},clip]{results/Joined_Illustration_Minimal.pdf}
% \end{graphicalabstract}
\end{frontmatter}
\twocolumn
\section{Introduction}
\begin{figure*}
\centering
\includegraphics[width=0.98\textwidth, trim={0 0 2cm 0},clip]{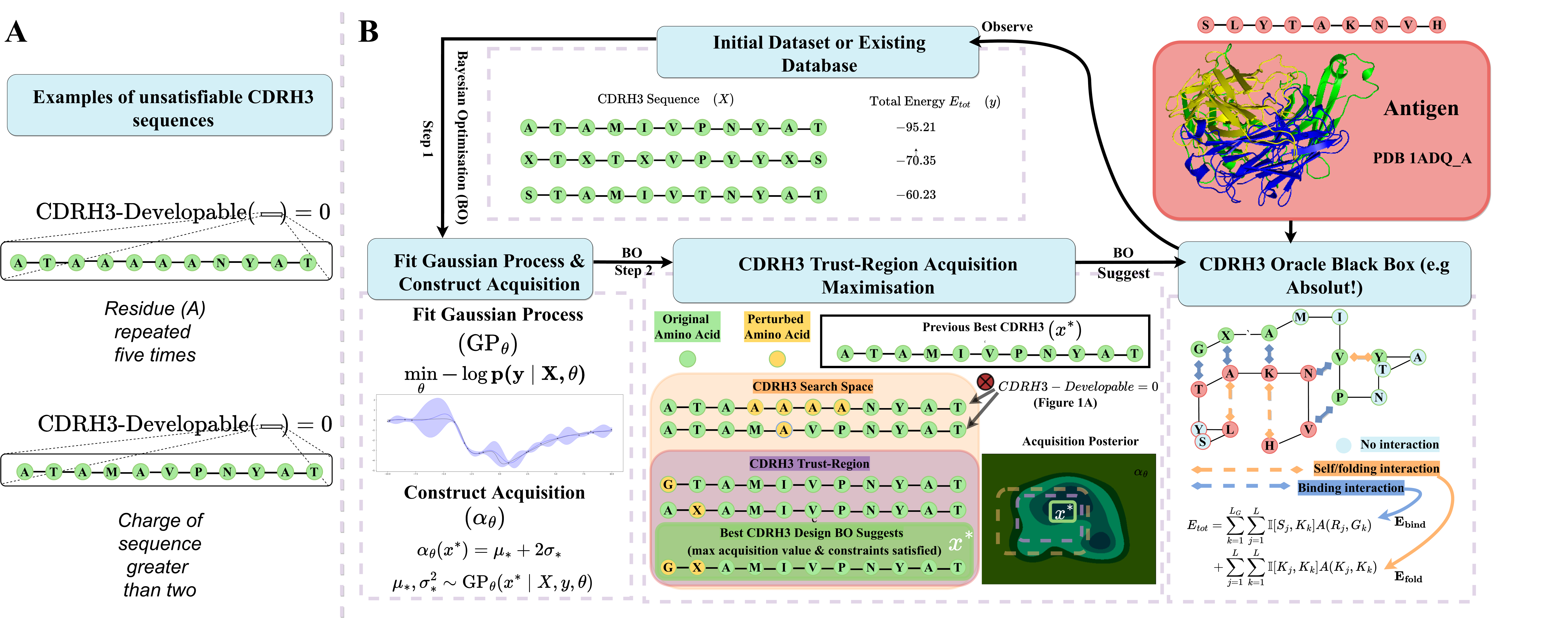}
     \caption{AntBO iteratively proposes a CDRH3 sequence and requests its affinity to Absolut!, before adapting its posterior with the affinity of this sequence. The performance of AntBO or other optimisation tools is measured as the highest affinity achieved and how fast it reaches high affinity. A. The demonstrative example of two CDRH3 sequences not satisfying the developability criterion is discarded in the overall optimisation procedure. B. overall optimisation process of \texttt{AntBO} for Antibody Design: from a predefined target antigen structure (discretised from its known PDB structure), binding affinities of antibody CDRH3 sequences to the antigen are simulated using Absolut!, as an in silico surrogate for costly experimental measurements. Absolut! is used as a black-box function to be optimised for $\EBRV_{\text{bind}}$ that is high-affinity CDRH3 protein designs within a trust region of acceptable sequences. }
    \label{fig:AntDesignIntro}
\end{figure*}
Antibodies or immunoglobulins (Igs) are utilised by the immune system to detect, bind and neutralise invading pathogens~\citep{Punt_2018}. From a structural perspective, these are mainly large Y-shaped proteins that contain variable regions, enabling specific molecular recognition of a broad range of molecular surfaces of foreign proteins called antigens ~\citep{Chothia1987,rajewsky1987evolutionary, xu2000diversity, akbar2021compact}. As a result, antibodies are a rapidly growing class of biotherapeutics~\citep{nelson2010development}. Monoclonal antibodies now constitute five of the ten top-selling drugs \citep{walsh2003biopharmaceutical,kaplon2018antibodies, Urquhart2021}. Antibodies are also utilised as affinity reagents in molecular biology research due to their ability to detect low concentrations of target antigens with high sensitivity and specificity~\citep{sela2013structural}.

A typical antibody structure consists of four protein domains: two heavy and two light chains connected by disulfide bonds. Each heavy chain (VH) includes three constant domains and one variable domain (Fv region), while a light chain (VL) possesses one constant and one variable domain~\citep{rajewsky1987evolutionary, xu2000diversity, rees2020understanding}. Antibodies selectively bind antigens through the tip of their variable regions, called the Fab domain (antigen-binding fragment), containing six loops, three on the light and three on the heavy chain, called complementarity-determining regions (CDRs) ~\citep{xu2000diversity, kunik2012paratome, robert2021one}. The interacting residues at the binding site between antibody and antigen are called the paratope on the antibody side and the epitope on the antigen side~\citep{xu2000diversity, kunik2012paratome, robert2021one}. {The base of an antibody is called the fragment crystallisable (Fc) region that reacts with the Fv region. Despite many studies focusing their attention only on Fv regions of antibodies and CDRH3 loops, in particular, it has been shown that the Fc region is also important for antibody design. The Fc region is connected to developability parameters such as aggregation, half-life, and stability which are crucial for antibody success in clinical trials~\cite{akbar2021progress}.}

The main overarching goal in computational antibody design is to develop CDR regions that bind to selected antigens (such as pathogens, tumour neoantigens, or therapeutic pathway targets) since the CDR regions mainly define the binding specificity ~\citep{cohn1980immunology,rajewsky1987evolutionary,norman2020computational}.  In particular, the CDRH3 region possesses the highest sequence and structural diversity, conferring a crucial role in forming the binding site~\citep{Chothia1987,xu2000diversity,akbar2021compact}. For this reason, the highly diverse CDRH3 is the most extensively re-engineered component in monoclonal antibody development. In this paper, we refer to the design of the CDRH3 region as an antibody design. 

When a candidate antibody-antigen complex structure is already known, structural methods predicting affinity change upon mutation at the interaction site ~\citep{morea2000antibody,clark2006affinity, clark2009antibody, nimrod2018computational} are useful in generating antibodies with higher affinity. As recent examples,~\citep{lippow2007computational} {combine structural} modelling and affinity scoring function to get a 140-fold affinity improvement on an anti-lysozyme antibody. {In contrast to other affinity-based scoring functions, ~\cite{kurumida2020predicting} use an ensemble ML strategy that utilises the affinity change induced by single-point mutations to predict new sequences with improved affinity. mCSM-AB2~\citep{myung2020mcsm} uses graph-based signatures to incorporate structural information of antibody-antigen complexes and combine it with energy inference using FoldX ~\citep{schymkowitz2005foldx} to predict improvements in binding energy.} Finally, two other generalised methods derived from the protein-protein interaction problem have been used on antibody affinity prediction: TopNetTree ~\citep{wang2020topology} combines a CNN with gradient-boosting trees, and GeoPPI ~\citep{liu2021deep} uses a graph neural network instead of the CNN. However, there is still a high discrepancy between the results of affinity prediction methods ~\citep{guest2021expanded, ambrosetti2020proabc}.

In practice, the development of antibodies is a complex process that requires various tools for building a structural model for different parts of the antibody~\citep{leem2016abodybuilder}, generating structures from antigen sequences~\citep{compiani2013computational}, and docking them~\citep{rawat2021exploring}. Moreover, the combinatorial nature of all possible CDRH3 sequences makes it impractical to query any antigen-antibody simulation framework exhaustively. For a sequence of length $L$ consisting of naturally occurring amino acids (AAs) ($N=20$), there are $N^L$ possible sequences. Thus, even with a modest size of $L=11$, this number becomes too large to search exhaustively. In reality, the search space is even larger since CDR sequence lengths can be up to $36$ residues~\citep{branden2012introduction}, and designed proteins are not restricted to naturally occurring AAs~\citep{2019_Yang}. Furthermore, not all CDRH3 sequences are of therapeutic interest. A CDRH3 can have a strong binding affinity to a specific target but may cause problems in manufacturing due to its unstable structure or show toxicity to the patient. Antibodies should be evaluated against typical properties known as developability scores for such reasons~\citep{akbar2022progress}. These scores measure properties of interest, such as whether a CDRH3 sequence is free of undesirable glycosylation motifs or the net charge of a sequence is in a prespecified range \citep{raybould2019five,bailly2020predicting}. 

Recently,~\citep{robert2021one} proposed \texttt{Absolut!}, a computational framework for generating antibody-antigen binding datasets that has been used to stress-test and benchmark different ML strategies for antibody-antigen binding prediction~\citep{robert2021one}. \texttt{Absolut!} is a deterministic tool that provides an end-to-end simulation of antibody-antigen binding affinity using coarse-grained lattice representations of proteins. We can use \texttt{Absolut!} to evaluate all possible binding conformations of an arbitrary CDRH3 sequence to an antigen of interest and return the optimal binding conformation. To be of real-world relevance, Absolut! preserves more than eight levels of biological complexity present in experimental datasets~\citep{robert2021one}: antigen topology, antigen AA composition, physiological CDRH3 sequences, a vast combinatorial space of possible binding conformations, positional AA dependencies in high-affinity sequences, a hierarchy of antigen regions with different immunogenicity levels, the complexity of paratope-epitope structural compatibility, and a functional binding landscape that is not well described by CDRH3 sequence similarity. Moreover, \texttt{Absolut!} demonstrates three examples where different ML strategies showed the same ranking in their performance compared with experimental datasets. Importantly, machine learning conclusions reached on Absolut!-generated simulated data transfer to real-world data \citep{robert2021one}. However, the combinatorial explosion of CDRH3 sequence space makes it unrealistic to exhaustively test every possible sequence, either experimentally or using Absolut!. Therefore, the problem of antibody-antigen binding design demands a sample-efficient solution to generate the CDRH3 region that binds an arbitrary antigen of interest while respecting developability constraints. 

Bayesian Optimisation (BO)~\citep{betro1991bayesian,mockus1978application, jones1998efficient,brochu2010tutorial} offers powerful machinery for aforementioned issues. BO uses Gaussian Processes (GP)~\citep{rasmussen2003gaussian} as a surrogate model of a black-box that incorporates the prior belief about the domain in guiding the search in the sequence space. The uncertainty quantification of GPs allows the acquisition maximisation to trade-off exploration and exploitation in the search space.\footnote{The acquisition maximisation step in the BO allows the tradeoff between exploration and exploitation. The idea of exploration is to eliminate the region of search space that does not contain the optimal solution with a high probability. The exploitation guarantees the search finds optimal sample points with a high probability. BO uses GP as a surrogate model that introduces mean and variance estimates with every data point. As BO encounters new data points in a local search to maximise the acquisition function, it checks if two points have the exact mean estimate and select the one with the highest variance, thereby exploring the space. When data points have the same variance, it chooses the one with the highest mean, thus exploitation.} This attractive property of BO enables us to develop a sample-efficient solution for antibody design. In this paper, we introduce \texttt{AntBO}---a combinatorial BO framework, for \textit{in silico} design of a target-specific antibody CDRH3 region. Our framework uses \texttt{Absolut!} binding energy simulator as a black-box oracle.  In principle, \texttt{AntBO} can be applied to any sequence region. Here, we consider the CDRH3 since this is the primary region of interest for antibody engineering~\citep{xu2000diversity,akbar2021compact,mason2021optimization,bachas2022antibody}. In addition, the Absolut! framework currently only allows CDRH3 binding simulation

Our key contributions are,
\begin{itemize}
\item The \texttt{AntBO} framework utilises biophysical properties of CDRH3 sequences as constraints in the combinatorial sequence space to facilitate the search for antibodies suitable for therapeutic development.
\item  We demonstrate the application of \texttt{AntBO} on $159$ known antigens of therapeutic interest. Our results demonstrate the benefits of \texttt{AntBO} for \textit{in silico} antibody design through diverse developability scores of discovered protein sequences.
\item \texttt{AntBO} substantially outperforms the very high-affinity sequences available out of a database of 6.9 million experimentally obtained CDRH3s, with several orders of magnitude fewer protein designs.
\item Considering the enormous costs (time and resources) of wet-lab antibody design-related experimentation, \texttt{AntBO} can suggest very high-affinity antibodies while making the fewest queries to a black-box oracle for affinity determination. This result serves as a proof of concept that \texttt{AntBO} can be deployed in the real world where sample efficiency is vital.
\end{itemize}
\section{Results}
\subsection{Formulating antibody design as a black-box optimisation with CDRH3 developability constraints}
\label{sec:bboxopt}
To design antibodies of therapeutic interest, we want to search for CDRH3 sequences with a high affinity towards the antigen of interest that satisfies specific biophysical properties, making them ideal for practical applications (e.g., manufacturing, improved shelf life, higher concentration doses). These properties are characterised as ``developability scores"~\citep{raybould2019five}. In this work, we use the three most relevant scores identified for CDRH3 region~\citep{raybould2019five,jin2021iterative}. First, the net charge of a sequence should be in the range $[-2,2]$. It is specified as a sum of the charge of individual residues in a primary amino-acid sequence. Consider a sequence $\xBRV=\{x_1,\cdots,x_n\}$, and let $\sI[.]$ the indicator function that takes value 1 if the conditions are satisfied and 0 otherwise, then the charge of a residue is defined as $C(x_i) = \sI [x_i\in \{R, K\}] + 0.1 \cdot\sI [x_i=H] - \sI [x_i\in \{D, E\}]$, and that of the sequence as $\sum_i C(x_i)$, where R stands for Arginine, K for Lysine, H for Histidine, D for Aspartic acid and E for Glutamate. Second, any residue should not repeat more than five times in a sequence $\sI[\text{count}(x_i)\leq 5 \mid \forall i\in [0,n-1]]$. Lastly, a sequence should not contain a glycosylation motif -- a subsequence of form N-X-S/T except when X is a Proline.

The binding affinity of an antibody and an antigen simulated as an energy score comes with several challenges. The energy score is not directly accessible as a closed-form expression that can return binding energy as a function of an input sequence without enumerating all possible binding structures. A vast space of CDRH3 sequences makes it computationally impractical to exhaustively search for an optimal sequence. Therefore, we pose the design of the CDRH3 region of antibodies as a black-box optimisation problem. Specific to our work, a black box refers to a tool that can take an arbitrary CDRH3 sequence as an input and return an energy score that describes its binding affinity towards a prespecified antigen. The high cost of lab experiments expects the antibody design method to suggest a sequence of interest in the fewest design steps. To simulate such a scenario, we want a sample-efficient solution that makes a very small prespecified number of calls to an oracle and suggests antibody sequences with very high affinity. 

To formally introduce the problem, consider the combinatorial space $\gX$ of protein sequences of length $n$, for $20$ unique amino acids, the cardinality of space is $|\gX|=20^n$. We can consider a black-box function $f$ as a mapping from protein sequences to a real-valued antigen specificity $f:\gX \rightarrow \sR$ where an optimum protein sequence under developability constraints is defined as,
\begin{align}
    \xBRV^{\ast} &= \arg\min_{\xBRV\in\gX} f(\xBRV),\nonumber\\
            &\text{s.t.} \quad \textbf{CDRH3-Developable}(\xBRV)
\end{align}
where $\textbf{CDRH3-Developable}:\gX \rightarrow \{0,1\}$ is a function that takes a sequence of amino acids and returns a Boolean value for whether constraints introduced in Section~\ref{sec:bboxopt} are satisfied (1) or unsatisfied (0). An example of unsatisfied CDRH3 sequence is shown in Figure~\ref{fig:AntDesignIntro} C.
\subsection{Combinatorial BO for antibody design}
Our goal is to search for an instance (antibody sequence) in the input space $\xBRV^{\ast}\in\gX$ that achieves an optimum value under the black-box function $f$. In a typical setting, the function $f$ has properties such as a) high evaluation cost, b) no analytical solution, and c) may not be differentiable. To circumvent these issues, we use BO to solve the optimisation problem. BO typically goes through the following loop; we first fit a GP on a random set of data points at the start. Next, optimise an acquisition function that utilises the GP posterior to propose new samples that improve over previous observations. At last, these new samples are added to data points to refit a GP and repeat the acquisition maximisation, as shown in Figure~\ref{fig:AntDesignIntro}. We have provided a brief introduction to BO in Supplementary~\ref{methods_implementation}. For a comprehensive overview of BO, we refer readers to~\citep{snoek2012practical, shahriari2015taking, hernandez2016general, frazier2018tutorial, cowen2020empirical,grosnit2021lsbo, garnett_bayesoptbook_2022}. 
\begin{figure*}[h!]
\centering
\includegraphics[width=.96\linewidth, trim={0 0cm 0 0},clip]{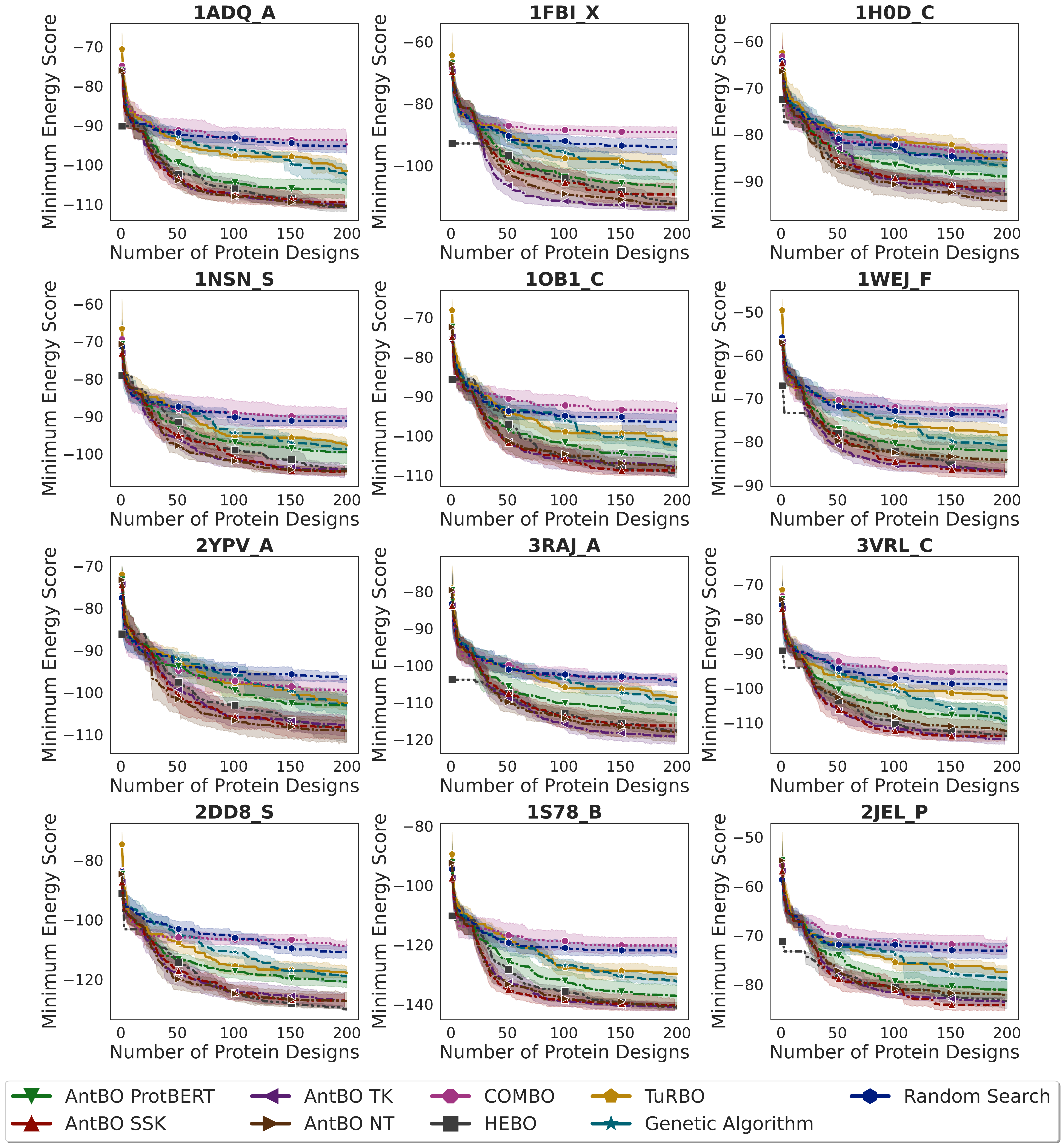}
\caption{AntBO is a sample-efficient solution for antibody design compared to existing baseline methods. AntBO with the transformed-overlap kernel can find binding antibodies whilst outperforming other methods. It takes around $38$ steps to suggest an antibody sequence that surpasses a very high-affinity sequence from~\texttt{Absolut! 6.9M} database and about $100$ to outperform a super+ affinity sequence. We run all methods with $10$ random seeds and report
the mean and $95\%$ confidence interval for the $12$ antigens of interest~\citep{robert2021one}. The title of each plot is a protein data bank (PDB) id followed by the chain of an antigen. The name of the disease associated with the antigen is provided in Table~\ref{tab:antboCDRH3s}. For extended results on the remaining $147$ antigens, we refer readers to~\ref{apsec:results}.
}
\label{fig:combinatorialeval}
\end{figure*}
%\paragraph{Kernels}
\subsubsection{Kernels to operate over antibody sequences}
\label{sec:kernel}
To build a GP surrogate model, we need a kernel function to measure a correlation between pairs of inputs. Since, in our problem, the input space is categorical, we need a kernel that can operate on sequences. We investigate three choices of kernels. Firstly, a transformed overlap kernel (TK) that uses hamming distance with a lengthscale hyperparameter for each dimension. Secondly, a protein BERT Kernel (protBERT) that uses pre-trained BERT model~\citep{brandes2021proteinbert} to map sequence to a continuous euclidean space and uses RBF kernel to measure correlation. Lastly, fast string kernel (SSK) defines the similarity between two sequences by measuring a number of common sub-strings of order $l$. The details are discussed in Methods~\ref{kernels}.
\subsubsection{CDRH3 trust region acquisition maximisation}\label{sec:cdrh3trustcombo}
The combinatorial explosion of antibody design space makes it impractical to use standard methods of acquisition maximisation. Several recent developments have proposed to use discrete optimisation algorithms for the combinatorial nature of problem~\citep{baptista2018bayesian, moss2020boss,buathong2020kernels,dadkhahi2020combinatorial,swersky2020amortized}. However, their application to antibody design requires a mechanism to restrict the search to sequences with feasible biophysical properties. We next introduce our novel method that utilises crucial biophysical properties to construct a trust region in the combinatorial sequence space. Thus, allowing us to extend the combinatorial BO machinery to antibody design.

At each iteration $t$ of the search step we define a trust region \texttt{CDRH3-TR} around the previous best point $\xBRV^{\ast}$ that includes all points satisfying antibody design constraints introduced in Section~\ref{sec:bboxopt} and differ in at most $L_t$ indices from $\xBRV^{\ast}$. We then run \texttt{CDRH3-TR} acquisition maximisation,
\begin{align}
    \label{eq:trustregion} 
    \texttt{CDRH3-TR}_{L_t}(\xBRV^{\ast}) &= \{\xBRV \mid \textbf{CDRH3-Developable}(\xBRV),\nonumber\\ 
    &\qquad\sum_{i}\delta(\xBRV_i, \xBRV^{\ast}_{i})\leq L_t
    \} 
\end{align}
where $\delta(.,.)$ is the Kronecker delta function.
To perform a search we start with the previous best $\xBRV^\ast$, next, sample a neighbour point $\xBRV^\ast_{\text{Neigh.}}$ contained within \texttt{CDRH3-TR}, by selecting a random amino acid and perturbing it with a new amino acid. We store the sequence if it improves upon the previous suggestions. The value of $L_t$ is restricted in the range $[d_{\min},d_{\max}]$, where $d_{\min}$ and $d_{\max}$ are the minimum and maximum size of TR that we treat as a hyperparameter. When $L_t$ reaches $d_{\min}$ we restart the optimisation using GP-UCB principle~\citep{srinivas2009gaussian}.It has been noted in several works~\citep{shylo2011restart, wan2021think} introducing $L_t$ promises theoretical convergence guarantees. Figure~\ref{fig:AntDesignIntro} illustrates this process. A detailed algorithm is presented in the Methods~\ref{alg:constrainedBO}.
\begin{table*}[ht!]
\setlength\tabcolsep{1.75pt}
    \centering
    \scalebox{0.91}{\begin{tabular}{l|rrr|rrr|rrr|rrr|rrr}
\toprule
            \bfseries &   \multicolumn{3}{c}{\bfseries Low} & \multicolumn{3}{c}{\bfseries  High} & \multicolumn{3}{c}{\bfseries Very High}& \multicolumn{3}{c}{\bfseries Super}& \multicolumn{3}{c}{\bfseries Super+}\\
             \bfseries \diagbox[innerwidth=3.25cm,font=\footnotesize]{Method}{Affinity} & \multicolumn{3}{c}{Top 5\%} & \multicolumn{3}{c}{Top 1\%} & \multicolumn{3}{c}{Top .1\%} & \multicolumn{3}{c}{Top .01\%} & \multicolumn{3}{c}{Best}\\
             &\# $\downarrow$ &\% $\uparrow$ &Score $\downarrow$ &\# $\downarrow$ &\% $\uparrow$ &Score $\downarrow$ &\# $\downarrow$ &\% $\uparrow$ &Score $\downarrow$ &\# $\downarrow$ &\% $\uparrow$ &Score $\downarrow$ &\# $\downarrow$ &\% $\uparrow$ &Score $\downarrow$
             \\
\midrule
AntBO TK & $20$ & $100$ & $0.2$ & $29$ & $100$ & $0.29$ & {$\boldsymbol{41}$} & {$\boldsymbol{99}$} & {$\boldsymbol{0.42}$} & {$\boldsymbol{58}$} & {$\boldsymbol{96}$} & {$\boldsymbol{0.6}$} & {$\boldsymbol{97}$} & {$\boldsymbol{55}$} & {$\boldsymbol{1.76}$} \\
{AntBO SSK} & {$21$} & {$100$} & {$0.21$} & {$30$} & {$100$} & {$0.3$} & {$46$} & {$100$} & {$0.46$} & {$64$} & {$96$} & {$0.67$} & {$94$} & {$48$} & {$1.94$} \\
{AntBO ProtBERT} & {$24$} & {$100$} & {$0.24$} & {$37$} & {$97$} & {$0.39$} & {$60$} & {$88$} & {$0.69$} & {$84$} & {$73$} & {$1.14$} & {$121$} & {$24$} & {$5.02$} \\
{AntBO NT} & {$19$} & {$100$} & {$0.19$} & {$28$} & {$99$} & {$0.29$} & {$43$} & {$98$} & {$0.44$} & {$61$} & {$95$} & {$0.64$} & {$111$} & {$52$} & {$2.15$} \\
{COMBO} & {$44$} & {$92$} & {$0.48$} & {$56$} & {$56$} & {$0.97$} & {$67$} & {$15$} & {$4.44$} & {$99$} & {$3$} & {$39.47$} & - & - & - \\
{HEBO} & {$\boldsymbol{15}$} & {$\boldsymbol{100}$} & {$\boldsymbol{0.15}$} & {$\boldsymbol{25}$} & {$\boldsymbol{100}$} & {$\boldsymbol{0.25}$} & {$50$} & {$100$} & {$0.5$} & {$74$} & {$97$} & {$0.75$} & {$130$} & {$56$} & {$2.33$} \\
{TuRBO} & {$34$} & {$100$} & {$0.34$} & {$65$} & {$99$} & {$0.65$} & {$109$} & {$79$} & {$1.37$} & {$124$} & {$40$} & {$3.11$} & {$112$} & {$1$} & {$134.4$} \\
{Genetic Algorithm} & {$34$} & {$100$} & {$0.34$} & {$70$} & {$99$} & {$0.7$} & {$111$} & {$92$} & {$1.21$} & {$140$} & {$56$} & {$2.48$} & {$141$} & {$4$} & {$33.79$} \\
{Random Search} & {$37$} & {$100$} & {$0.37$} & {$71$} & {$78$} & {$0.9$} & {$84$} & {$23$} & {$3.61$} & {$99$} & {$3$} & {$39.6$} & - & - & - \\
{LamBO} & {$19$} & {$100$} & {$0.19$} & {$32$} & {$100$} & {$0.32$} & {$55$} & {$100$} & {$0.55$} & {$73$} & {$94$} & {$0.78$} & {$101$} & {$51$} & {$1.99$} \\
\bottomrule
\end{tabular}}
    \caption{AntBO consistently ranks as the best method in designing high-affinity binding antibodies while making minimum calls to the black box oracle. Here we analyse the required number of successful trials to reach various binding affinity categories. We report a number of protein designs needed to reach low, high, very high and super affinity (top 5\%, 1\%, 0.1\%, 0.01\% quantiles from \texttt{Absolut! 6.9M} database). We denote by super+ as the number of designs required to outperform the best CDRH3 in the 6.9M database. The various binding categories are taken from existing works~\citep{robert2021one,akbar2021silico}. We collectively report three scores for every affinity class across all respective methods (across $10$ trials and $12$ antigens). For a given method, let $\textbf{TE}$ be a matrix of size $[12\times10, 200]$ (where each trial lasts 200 iterations) of all trial affinities, $\sI(\textbf{TE}_i \leq c )$ be an indicator function which returns $1$ if for a given trial $i$ a method finds any $\textbf{TE}$ better than the affinity category $c$'s value, and $\mathcal{F}$ as a function which returns minimum samples required to reach affinity category $c$, if the trial did not reach the affinity category it returns $0$. \\
    The first column {(\# $\uparrow$)} outlines the average number of protein designs $\sum_i^N\mathcal{F}(\textbf{TE}_i , c )/ N$ required to reach the respective affinity quantile value $c$. The second column {(\% $\uparrow$)} is the proportion of trials $\sum_i^N \sI(\textbf{TE}_i \leq c )/N$ that output a protein design better than the given affinity category $c${, given as a percentage}. Ideally, the best method would attain the lowest value in the first column and a value of {$100$\%} in the second column, showing that it reaches the affinity category in ALL trials and does so in the lowest number of samples on average. Due to the importance of both measurements, in the third column {(Score $\downarrow$)}, we report the ratio of {the} two values to get an estimate of overall performance, where we penalise the reported mean samples required to reach an affinity category by the \% of failed trials to reach that affinity category. The penalised ratio balances the probability of designing a Super+ sequence and required evaluations. The categories in which no samples by a method reach the affinity class are denoted by $-$. Our results demonstrate that \texttt{AntBO TK} is the superior method that consistently takes fewer protein designs to reach important affinity categories.}
    \label{tab:bindingclass}
\end{table*}
\subsection{Evaluation setup and baseline methods}
\label{subsec:resultsmain}
We use \texttt{Absolut!} for simulating the energy of the antibody-antigen complex. We indicate our framework (\texttt{AntBO}'s) kernel choice directly in the label e.g. \texttt{AntBO SSK}, \texttt{AntBO TK}, \texttt{AntBO ProtBERT.} We compare \texttt{AntBO} with several other combinatorial black-box optimisation methods such as \texttt{HEBO}~\citep{cowen2020empirical}, \texttt{COMBO}~\citep{Oh_2019}, \texttt{TuRBO}~\citep{eriksson2019scalable}, {LamBO}~\citep{stanton2022accelerating}. random search (RS) and Genetic Algorithm (GA). We introduce the same developability criteria defined as CDRH3 trust region in all the methods for a fair comparison. {We also run \texttt{AntBO TK} without hamming distance criterion that is omitting $\sum_{i}\delta(\xBRV_i, \xBRV^{\ast}_{i})\leq L_t$ in the Equation~\ref{eq:trustregion} of trust-region and label it as \texttt{AntBO NT}. The trust region size $L_t$ is the distance of the best-seen sequence from the random starting sequence. The criterion restricts the local optimisation within a certain radius $L_t$ from the starting sequence. Removing the distance criterion $d_{\max}$ allows a local search to reach the maximum possible value and lets the optimisation process explore distant regions in the search space.} For an explanation of the algorithms, including the configuration of hyperparameters, we refer readers to Supplementary~\ref{subsec:baseline}. For the primary analysis in the main paper, we use twelve core antigens: their protein data bank (PDB) id, chain of antigen, and the associated disease name provided in Table~\ref{tab:antboCDRH3s} of Supplementary. Our choice of antigens is based on their interest in several studies~\citep{robert2021one, akbar2021silico}. We also evaluate our approach on the remaining $147$ antigens in \texttt{Absolut!} antibody-antigen binding database. Results for those are provided in~\ref{apsec:results}.

\begin{figure*}[t]
    \centering
    \includegraphics[clip, trim=11cm 3.5cm 9cm 3.5cm, width=\textwidth]{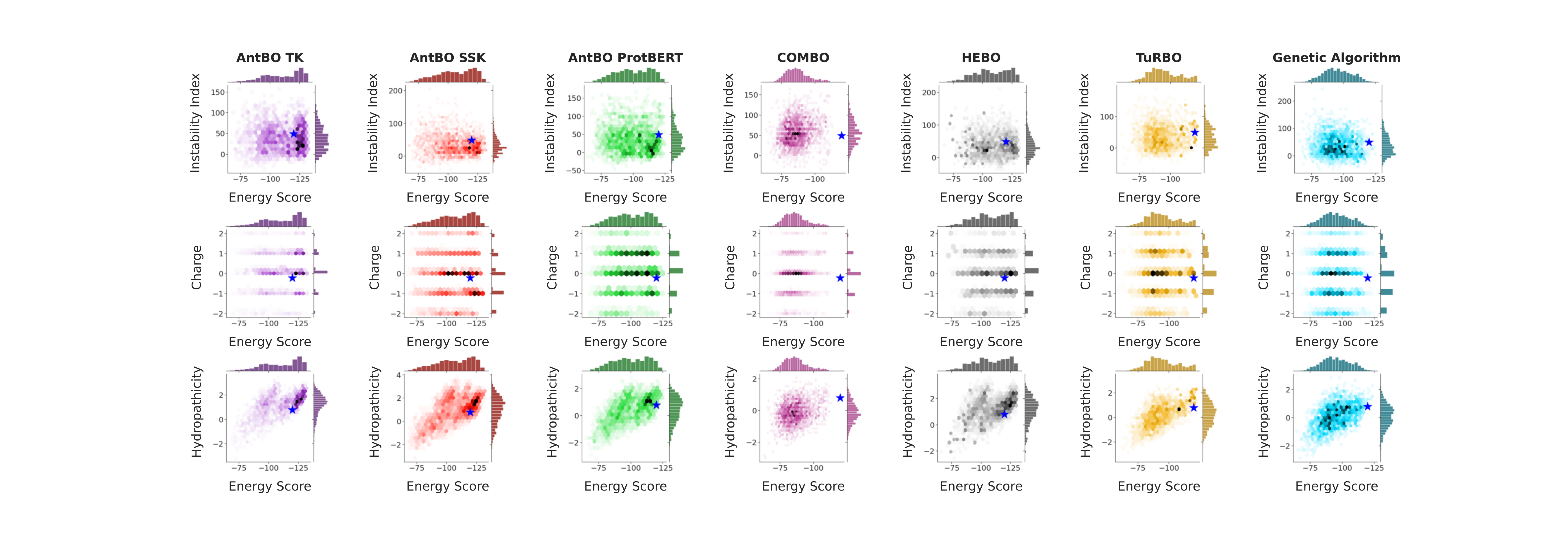}
    \caption{AntBO can design antibodies that achieve diverse developability scores, demonstrating it is a viable method to be practically investigated. We analyse the developability scores of  $200$ proteins designed by each method {averaged across all $10$ random seeds} to simulate the diversity of suggested proteins across a single trial. Here, we report developability scores for S protein from the SARS-CoV virus (PDB id: 2DD8). The landscape of designed sequences suggested during the optimisation process for each method is shown with their binding affinity and three developability scores (Hydropathicity, charge and instability). {We also take Super+ (top 0.01\%) sequences from \texttt{Absolut!} 6.9M database and report their mean developability scores denoted by $\star$ in the plots.} Interestingly, we observe a positive correlation between Hydropathicity increasing with energy. While other methods have a larger charge spread, we see \texttt{AntBO} favourably suggesting the most points with a neutral charge. {We observe the spread of developability scores of \texttt{AntBO} methods is close to the average score of Super+ sequences.} Overall, we conclude that energetically favourable sequences still explore a diverse range of developability scores and that the protein designs of \texttt{AntBO} are more stable than other methods. }
    \label{fig:dev_scores}
\end{figure*}
\subsection{\texttt{AntBO} is {sample-efficient} compared to baseline methods}
\begin{figure*}[t]
    \centering
    \includegraphics[width=\textwidth]{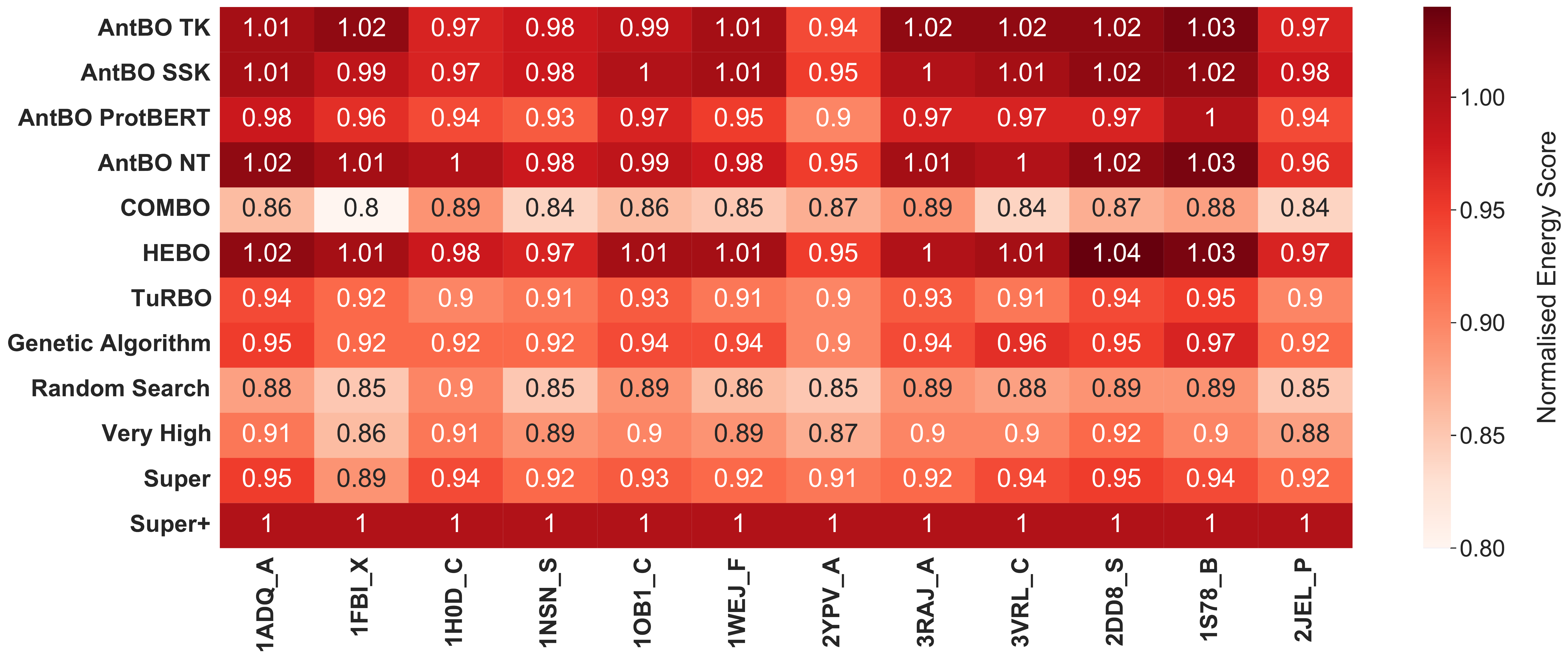}
    \caption{{We compare the binding energy threshold of different categories (Low, High, Very High, Super, Super+) obtained from Absolut! 6.9M database and the average binding affinity of a sequence designed using \texttt{AntBO} methods and the baselines. The energy scores are normalised by the threshold of the Super+ category. We observe \texttt{AntBO} outperforms the best sequence in a majority of antigens and emerges as the best method in finding high binding affinity sequences in under $200$ evaluations.}}
    \label{fig:bindingcategories}
\end{figure*}

Precise wet-lab evaluation of an antibody is a tedious process and comes with a significant experimental burden because it requires purifying {both antibody and antigen and testing their binding} affinity~\citep{rawat2021exploring, laustsen2021animal}. We, therefore, first investigate the sample efficiency of all optimisation methods. We ran experiments with a pre-specified budget of $200$ function calls and reported the convergence curve of protein designs vs minimum energy (or binding affinity) in Figure~\ref{fig:combinatorialeval}. {The experimental validation we substitute here by Absolut!-based in silico proof is expensive and time-consuming. Therefore, budgeting of optimisation steps is a vital constraint~\citep{laustsen2021animal}.}
{In Figure~\ref{fig:bindingcategories}, we compare AntBO with baseline methods and various binding affinity categories \texttt{Very High}, \texttt{Super} and \texttt{Super+} (determined from 6.9 million (6.9M) experimentally obtained murine CDRH3s available from \texttt{Absolut!} database). We normalise the energy score by the \texttt{Super+} threshold}. Core antigen experiments are run with ten random seeds and the remaining antigens with three. We report the mean and $95\%$ confidence interval of the results.

%st{We also include horizontal lines showing \texttt{Very High}, \texttt{Super} and \texttt{Super+} Affinity. st{We remind the reader that} \texttt{Super+} is the best energy score out of the 6.9 million (6.9M) experimentally obtained murine CDRH3s available from \texttt{Absolut!}}

We observe \texttt{AntBO TK} achieves the best performance w.r.t to minimising energy (maximising affinity), typically reaching high affinity within $200$ protein designs, with no prior knowledge of the problem, \texttt{AntBO TK} can search for CDRH3 sequences that achieve significantly better affinity than very-high affinity sequences from experimentally obtained~\texttt{Abs-\linebreak olut! 6.9M} database. In majority of antigens \texttt{AntBO TK} outperforms the best evaluated CDRH3 sequence by \texttt{Absolut!}. We noticed that for the S protein chain of 1NSN, P protein chain of 2JEL, and on a few other antigens (results outlined in~\ref{apsec:results}) \texttt{AntBO} gets close to the best experimental sequence {known for that antigen} but doesn't outperform its affinity. We attribute this result to the complexity of the 3D lattice representation of an antigen that might require more sequence designs to explore the antibody optimisation landscape. We wish to study this effect in future work. {For some antigens such as 1H0D, the binding affinity decreases in smaller factors when compared to other antigens such as 1S78. This observation shows that some antigens are difficult to bind, while there are more possible improvements for others. ~\cite{akbar2022progress} make a similar observation where transfer learning from one antigen to another differed across different pairs.}

We found on majority of antigens \texttt{AntBO TK} outperforms \texttt{AntBO ProtBERT}. This finding contradicts our assumption that a transformer trained on millions of protein sequences would provide us with a continuous representation that can be a good inductive bias for GP. We believe this could be associated with specific characteristics of antibody sequences that differ from a large set of general protein sequences. Consequently, there is a shift in distribution between the sequences used for training the protein BERT model and the sequences we encounter in exploring the antibody landscape. This finding also demonstrates that AntBO can reliably search in combinatorial space without relying on deep learning models trained on enormous datasets. However, we want to remark that \texttt{AntBO ProtBERT} performs on par with other baselines.

We next investigate the average number of protein designs \texttt{AntBO TK} takes to get to various levels of binding affinity across all antigens. For this purpose, we take five affinity groups from existing works~\citep{robert2021one,akbar2021silico}: low affinity(5\%), high affinity(1\%), very high affinity (0.1\%), super (0.01\%) and super+ (the best known binding sequence taken from 6.9M database.) and report the average protein designs needed to suggest a sequence in the respective classes for $188$ antigens. Table~\ref{tab:bindingclass} describes the performance of \texttt{AntBO TK} and other baselines. We observe \texttt{AntBO TK} reaches a very high-affinity class in around $\sim 38$ protein designs, super in around $50$ designs and only $85$ to outperform the best available sequence. This sample efficiency of \texttt{AntBO TK} demonstrates its superiority and relevance in the practical world. 
%Due to a computational budget, we could not run all methods across all antigens; however, we report similar results in the Supplementary (Table~\ref{tab:bindingclasscore}) for all methods across the 12 core antigens.\\

%{ToDo: Discuss AntBO 3D plot @philippe}

% \begin{figure}
%     \centering
%     \includegraphics{results/StructuresStepsBy10/0-RRU.bmp}
%     \caption{Caption}
%     \label{fig:my_label}
% \end{figure}

\subsection{\texttt{AntBO} suggests antibodies with favourable developability scores}
AntBO iteratively designs antibodies that improve (over previous suggestions) to reach an optimal binding sequence. The antibody sequences we encounter in the iterative refinement process compose a trajectory on the binding affinity landscape. To understand the search mechanism of AntBO, here, we investigate the developability scores of $200$ CDRH3 designs found along the above trajectory on the binding affinity landscape. This analysis helps us understand how optimising energy affects the biophysical properties of antibody sequences. The developability scores we used in the \texttt{CDRH3-TR} are a few of many other biophysical properties. As noted in Section~\ref{sec:bboxopt}, more scores can be added as constraints. However, finding an optimum sequence also adds an extra computational cost. Here, in addition to charge, we report hydrophobicity (HP) and instability index, which have been used in other studies for assessing downstream risks of antibodies~\citep{robert2021one,mason2021optimization}. A smaller instability index value means the sequence has high conformational stability, and in practical scenarios, it is desired to have a score of less than $40$. CDRH3 regions tend to aggregate when developing antibodies, making it impractical to design them. This phenomenon is due to the presence of hydrophobic regions. A low value of HP means a sequence has a lesser tendency to aggregate. We use the Biopython~\citep{chapman2000biopython} package to compute HP and instability scores. We next discuss the analysis of developability parameters for severe acute respiratory syndrome coronavirus (SARS-CoV) antigen. The results on the remaining core antigens are provided in~\ref{apsec:results}.
\subsection{Case Study: Application of \texttt{AntBO} for SARS-CoV antibody design}
\label{subsec:resultsSARS}
The S protein of the SARS-CoV (PDB id: 2DD8) is responsible for the entry of the virus into the host cell, making it an important therapeutic target for the effective neutralisation of the virus. Figure~\ref{fig:dev_scores} demonstrates that AntBO can design antibodies for SARS-CoV with diverse developability parameters. On the top of each plot is a histogram of binding affinity of $200$ designs and a right histogram of developability scores. The hexagon discretises the space with binding affinity on the vertical axis and developability score on the horizontal axis. The colour of hexagons shows a subspace frequency within a specific binding affinity range and the respective developability score.  We observe the distribution of three developability scores varies across all methods showing the distinction between their designed sequences. Interestingly, the performance on developability scores, which were not included in constraints, demonstrates that the \texttt{AntBO} methods can identify sequences with diverse developability parameters. This observation suggests that our approach is suitable for exploring sequences towards high affinity and selecting candidates in a desired developability region. {To understand how the spread of scores compares to experimentally known sequences. We take a set of super+ (top 0.01\% annotated using Absolut! 6.9M database) and report their average developability score denoted by $\star$ sign in the hexagram plots. We observe the spread of scores of AntBO methods is close to the mean of super+.} Thus, we can conclude \texttt{AntBO} is a more practically viable method for antibody design. 
%We also did a similar analysis for other core antigens. The results are described in~\ref{apsec:results}.
\begin{figure*}
\centering
    \includegraphics[width=1.9\columnwidth]{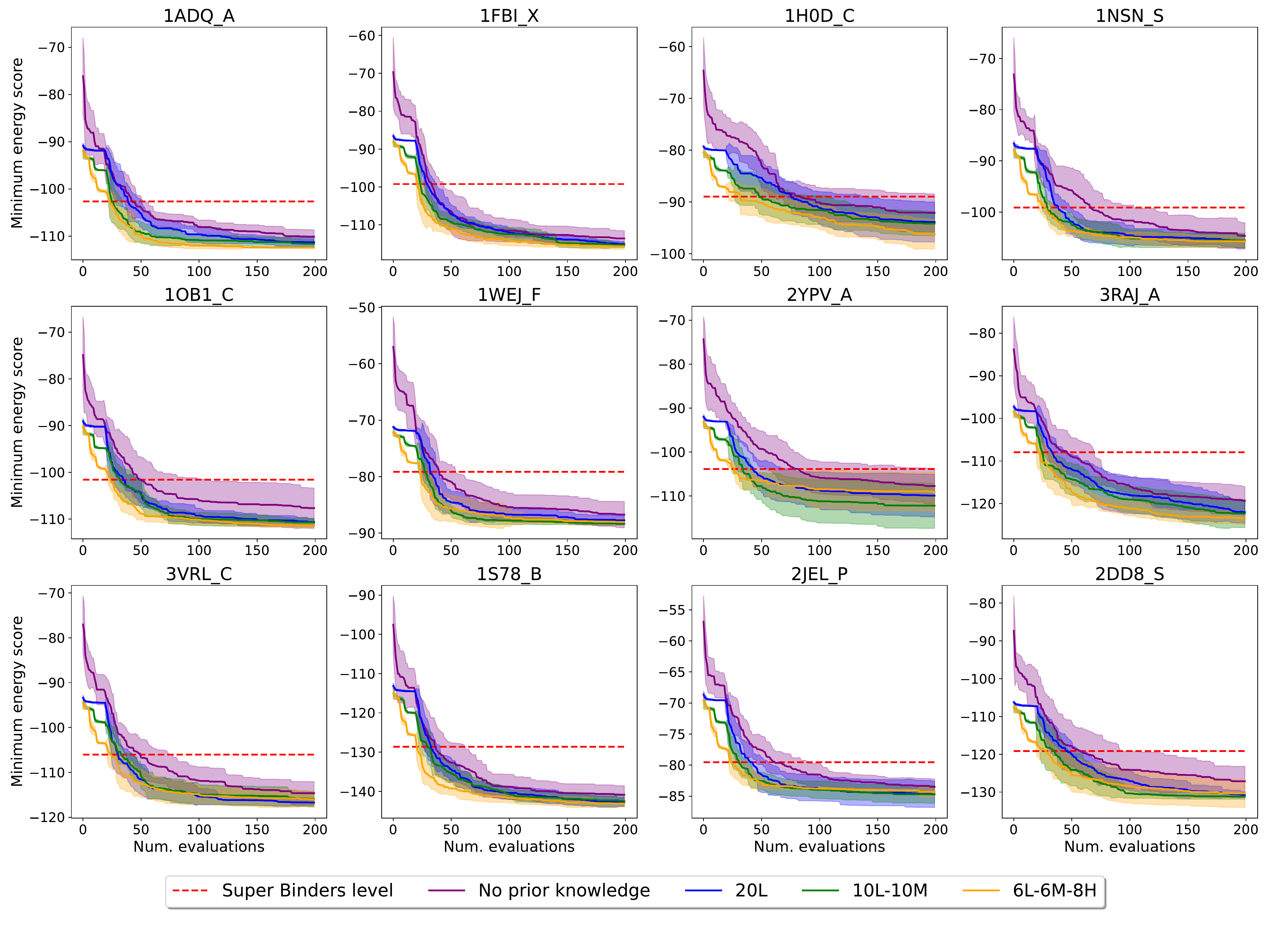}
    \vspace{-0.5cm}
    \includegraphics[width=1.9\columnwidth]{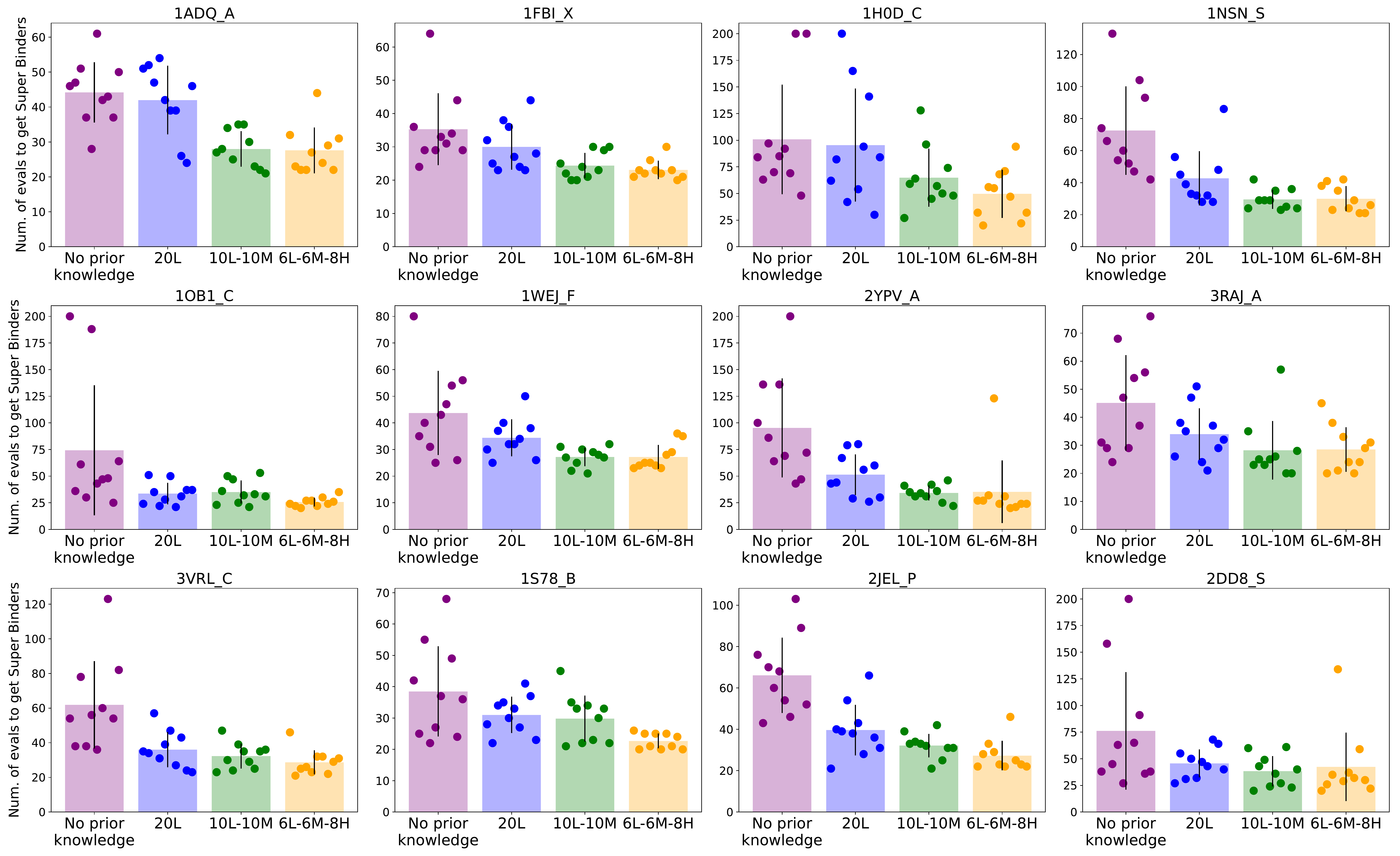}
    \caption{Effect of different initial class distributions on BO convergence. Experiments are run for three sets of initial points varying with the amount of binder (top 1\%) and non-binders (remaining sequences): losers 20L (with only non-binders),  mascotte 10L-10M (half non-binders and half low binders) and heroes 6L-6M-8H (six non-binders, six low-binders and eight high binders). {Top} is the BO convergence plot with a horizontal line denoting the energy threshold to reach the super binder level. {Bottom figures show} the histogram of the number of antibody designs required to reach super binding affinity class averaged across $5$ trials. We find for the majority of antigens, prior knowledge of binders helps in reducing the number of evaluations.}
    \label{fig:init_bo}
\end{figure*}
\subsection{Knowledge of existing binders benefits \texttt{AntBO} in reducing the number of calls to black-box oracle}
The optimisation process of \texttt{AntBO} starts with a random set of initial points used in fitting the GP surrogate model. This initialisation scheme includes the space of non-binders, allowing
more exploration of the antibody landscape. Alternatively, we can start with a known set of binding sequences to allow the surrogate model to better exploit the local region around the good points in finding an optimum binding sequence. We hypothesise that the choice of initial data points dictates the tradeoff between exploration and exploitation of the protein landscape. To investigate the question, we study the effect of different initialisation schemes on the number of function evaluations required to find very-high affinity sequences. We create three data points categories: losers, mascotte, and heroes. In the losers, all data points are non-binders; in the mascotte we use half non-binders and half low-binders, and finally, in the heroes, we take a proportion of six non-binders, six low-binders, and eight high-binders. The threshold of categories is obtained using \texttt{Absolut!} database~\citep{robert2021one}. 

For each of the $12$ core antigens, Figure~\ref{fig:init_bo} reports the convergence plot and the histogram of an average number of evaluations across $5$ trials required to reach the super affinity category. When starting with the known sequence, \texttt{AntBO} exploits the prior knowledge of the landscape, limiting the search technique to find an optimal design in the vicinity of available binders. Interestingly, we observe that using prior information of binders for some antigens such as 2YPV\textunderscore A, 3RAJ\textunderscore A, \texttt{AntBO} require more sequence designs to reach the super affinity category. We hypothesise this phenomenon can be attributed to the complexity of antigen structure that, in turn, can benefit from more exploration of the antibody sequence landscape. 
%In a situation with a more complex antigen structure we can benefit from the exploration of antibody landscape.

\begin{figure}
    \centering
    \includegraphics[width=0.9\columnwidth]{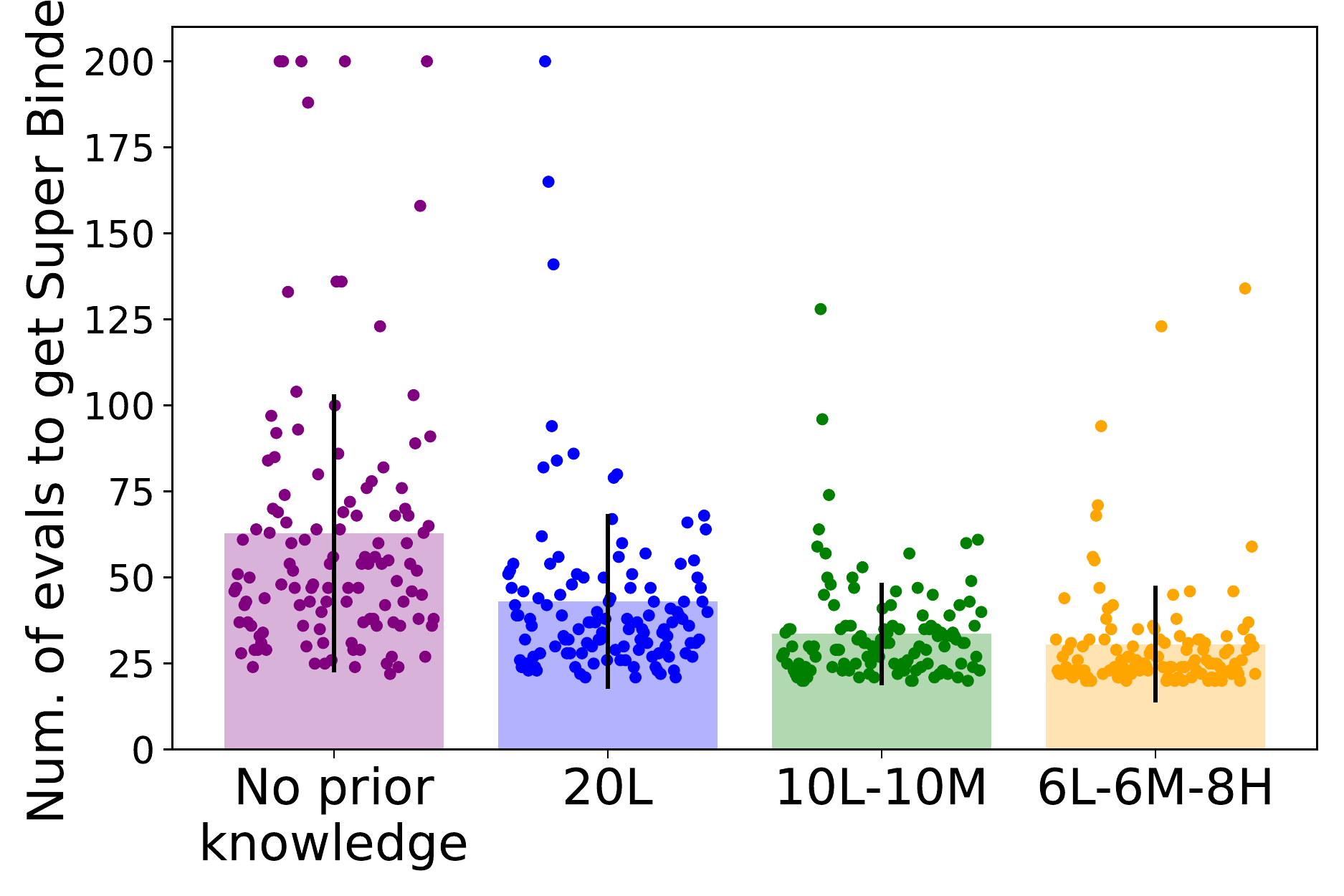}
    \caption{AntBO benefits from the knowledge of prior binding sequence in arriving at Super binders. The average number of antibody designs reduces when information about known binders is made available to GP surrogate model. On the y-axis, we report the average number of iterations required across all antigens to reach the Super binding affinity class (outperforming the best sequence in Absolut! database), and on the x-axis, we have three affinity classes, namely losers 20L (with only non-binders),  mascotte 10L-10M (half non-binders and half low binders) and heroes 6L-6M-8H (six non-binders, six low-binders and eight high binders)}
    \label{fig:init_bo_avg_hist}
\end{figure}

Figure~\ref{fig:init_bo_avg_hist} further reports the histogram of number of antibody design averaged across both $5$ trials and $12$ core antigens. We observe an overall required number of calls to the black-box oracle to reach the super binding affinity category decreases when information on known antibody binding sequences is made available to AntBO as training data for the GP surrogate model. We can interpret the initialisation as a prior domain knowledge that aids the antibody design process by reducing the computational cost of evaluating the black-box oracle.

\section{Discussion}
\subsection{General computational approaches for antibody discovery}
Several computational approaches have been developed to support antibody design ~\citep{norman2020computational, akbar2021progress}, either using physics-based antibody and antigen structure modelling ~\citep{fiser2003modeller,almagro2014second, leem2016abodybuilder} and docking ~\citep{brenke2012application,sircar2010snugdock}, or using machine learning methods to  learn the rules of antibody-antigen binding directly from sequence or structural datasets ~\citep{akbar2021progress}. (1) Paratope and epitope prediction tools consider either sequence or structure of both antigen or antibody to predict the interacting residues ~\citep{soria2015overview, lu2021structure, sela2015antibody, jespersen2019antibody, krawczyk2014improving, liberis2018parapred, ambrosetti2020proabc, kunik2012paratome, krawczyk2013antibody, del2021neural}. Knowledge of the paratope and epitope does not directly inform affinity but rather helps prioritise important residues to improve affinity. (2) Binding prediction tools, often inspired by Protein-Protein Interaction prediction (PPI) tools ~\citep{liu2018machine}, predict the compatibility between an antibody and an antigen sequence or structure. The compatibility criterion is decided by either using clustering to predict sequences that bind to the same target~\citep{wong2021ab, xu2019functional}, paratope-epitope prediction model ~\citep{akbar2021compact} or using a ranking of binding poses to classify binding sequences~\citep{schneider2022dlab}. However, predicting antibody binding mimics the experimental screening for antibody candidates but does not directly help to get high affinity and specific antibody sequences. (3) Affinity prediction tools specifically predict affinity improvement following mutations on antibody or antigen sequences. Our work particularly focuses on the affinity prediction problem because it is a major time and cost bottleneck in antibody design.
\subsection{Small size of available experimental datasets limiting the application of ML methods}
{\subsubsection{Available experimental datasets}}
The experimental datasets describing antibody binding landscape can be categorised in four ways : (1) Structures of antibody-antigen complexes provide the most accurate description of the binding mode of an antibody and the involved paratope and epitope residues ~\citep{schneider2022sabdab}, which helps to prioritise residues that can modulate binding affinity. Structures do not directly give an affinity measurement but can be leveraged with molecular docking and energy tools to infer approximate binding energy. Only $\sim1200$ non-redundant antibody-antigen complexes are known so far ~\citep{schneider2022sabdab}. (2) Sequence-based datasets contain the results of qualitative screenings of thousands of antibodies (either from manually generated sequence libraries or from ex vivo B cells) ~\citep{laustsen2021animal}. Typically, millions of sequences can be inserted into carrier cells expressing the antibody on their surface. Following repeated enrichment steps for binding to the target antigen, a few thousand ‘high affinity’ sequences can be obtained ~\citep{mason2021optimization}, and newer experimental platforms will soon allow reach a few million. As of yet, however, sequencing datasets can only label sequences with binder or non-binder; or low affinity, medium affinity, and high-affinity classes. (3) Affinity measurements are very time-consuming because they require the production of one particular antibody sequence as protein before measuring its physicochemical properties (including other \textit{in-vitro} measurable developability parameters). Affinity measurements are precise and quantitative, either giving an affinity reminiscent of the binding energy or down to an association and dissociation constant. As an example, the AB-bind database only reports in total ~1100 affinities on antibody variants targeting 25 antigens ~\citep{sirin2016ab}, and a recent cutting-edge study ~\citep{mason2021optimization} measured the affinity of 30 candidate antibodies, showing the experimental difficulty to obtain the affinity measurement of many antibodies. Finally, (4) \textit{in} (and \textit{ex}) \textit{vivo} experiments describe the activity of injected antibodies, including \textit{in vivo} developability parameters ~\citep{raybould2019five, xu2019structure} such as half-life, and toxicity including off-targets. \textit{In vivo} experiments are restricted to lead candidates due to their high cost and cannot be performed when screening for antibody leads. Although qualitative (sequencing) datasets inform on initial antibody candidates, increasing the activity and specificity of antibody candidates requires many steps to further improve their affinity towards the antigen target while keeping favourable developability parameters. It is the most tedious and time-consuming step. {While new-coming methods may reveal more quantitative affinity measurements at high throughput~\citep{adams2016measuring}.} 
{
\subsubsection{Generative models for sampling new antibody candidates} 
Generative ML architectures have been leveraged to generate antibody candidates from sequence datasets. Specifically, an autoregressive model~\citep{sutskever2011generating}, a variational autoencoder~\citep{kingma2013auto} or a generative adversarial network (GANs)~\citep{goodfellow2014generative} have been used for generating amino-acid sequences of antibodies~\citep{amimeur2020designing,eguchi2020ig, shin2021protein,akbar2021silico,shuai2021generative,leem2021deciphering}.  ~\citep{amimeur2020designing} also incorporate therapeutic constraints to avoid sampling a non-feasible sequence at inference.~\citep{ingraham2019generative,koga2012principles,cao2021fold2seq} additionally includes the information of a backbone structure. Recently ~\citep{jin2021iterative} proposed an iterative refinement approach to redesign the 3D structure and sequence of antibodies for improving properties such as neutralising score. The generative modelling paradigm can increase the efficient design of antibodies by prioritising the next candidates to be tested experimentally.
}

Due to the current small size of datasets, the application of ML methods for improving antibody affinity has been minimal. Further, the generalisability of such approaches is difficult to assess, and there is a lack of generative models that can be conditioned for affinity. Here, we set out to leverage the maximal information on antibody sequence affinity from the minimal number of experimental, iterative measurements using BO to generate an informed prediction on potential higher affinity sequences. We use the Absolut! simulator as a black-box oracle to provide a complex antibody-antigen landscape that recapitulates many layers of the experimental complexity of antibody-antigen binding. 
{\subsection{Combinatorial methods for protein engineering}Methods on protein engineering~\citep{romero2009exploring,goldsmith2017enzyme, zeymer2018directed} use evolutionary methods to explore the combinatorial space of protein sequences. They use directed evolution -- an iterative protocol of mutation and selection followed by a screening to identify sequences with improved diversity and functional properties. However, the approach suffers from high experimental costs due to inefficient screening methods. To overcome the experimental hurdle~\citep{2018_Yang} propose an ML pipeline for protein engineering. The central theme is to utilise the measurements of known protein sequences to train an ML model that can further guide the evolution of protein sequences. In a concurrent work~\citep{stanton2022accelerating} introduce LamBO -- a multi-objective BO framework for designing molecular sequences. LamBO utilises a deep kernel for fitting GP. Specifically, it does optimisation in the latent space of a denoising autoencoder. We want to remind the readers \texttt{AntBO} with protBERT uses a deep kernel in the latent space of pre-trained BERT for training GP. However, the acquisition maximisation is done in the input space. The major limitation here is that none of these methods has been investigated for antibody design due to limited data on antibody specificity.}
%Not only does \texttt{AntBO} confidently generate diverse new CDRH3 sequences with higher affinity than the previously known ones, but its interpretable architecture offers the possibility to extract the reasons for higher affinity in silico de novo generated sequences.
%\section{Conclusion}
%We have proposed \texttt{AntBO}, a combinatorial BO framework for designing CDRH3 regions of antibodies.
\subsection{\texttt{AntBO} a {sample-efficient} solution for computationally favorable antibody design}
A list of therapeutically relevant developability parameters is considered vital for designing antibodies~\citep{mason2021optimization, akbar2021progress}. These parameters include solubility, charge, aggregation, thermal stability, viscosity, immunogenicity (i.e., the antibody should not induce an immune response, which might also induce its faster clearance by the body), glycosylation motifs, and the \textit{in vivo} half-life. Although the whole antibody sequence can be modified to improve developability, the CDRH3 region seems also to have a critical impact on them beyond only affinity and antigen recognition ~\citep{grevysantibody}. Therefore, it is crucial to include developability constraints in CDRH3 design. Interestingly, many parameters can be calculated in advance from the antibody sequence according to experimentally validated estimators ~\citep{akbar2021progress}, allowing for defining boundaries of the search space according to development needs. Our proposed \texttt{AntBO} framework utilises the developability parameters to construct a trust region of feasible sequences in the combinatorial space, thus allowing us to search for antibodies with desired biophysical properties.

Our findings across several antigens demonstrate the efficiency of \texttt{AntBO} in finding sequences outperforming many baselines, including the best CDRH3 obtained from the \texttt{Absolut!} 6.9 million database. \texttt{AntBO} can suggest very high-affinity sequences with an average of only 38 protein designs and a super binding sequence within 100 designs. {The versatility of Absolut! allows defining binder/non-binder levels based on user requirements. In the future, an interesting investigation would be measuring the performance of \texttt{AntBO} as a function of different binder definitions~\citep{robert2021one}. We also wish to investigate our framework for improved structure prediction with other docking simulation models and perform experimental validations.}
% Since, \texttt{AntBO} sequentially designs one antibody per evaluation. To achieve a more efficient experimental scenario \texttt{AntBO} can be adapted to design a batch of sequences allowing us to draw even more sequences in fewer evaluations.We will also investigate the choice of other relevant criteria beyond energy by multi-objective BO. 
%For a structure prediction, we can use a realistic benchmark such as AbodyBuilder~\citep{leem2016abodybuilder}, AlphaFold Multimer~\citep{Evans2021.10.04.463034}, and for docking a simulator such as FoldX~\citep{schymkowitz2005foldx}. We will also investigate the choice of other relevant criteria beyond energy by multi-objective BO.
{
\section{Limitation}
We want to remark to the readers that \texttt{AntBO} is the first framework showcasing different flavours of combinatorial BO for the antibody design problem. The potential limitations of \texttt{AntBO} in its current scope are: i) \texttt{AntBO} sequentially designs antibodies suggesting one sequence per evaluation step. To achieve a more efficient experimental scenario, \texttt{AntBO} can be adapted to a batch scenario, allowing us to design more sequences in fewer evaluations. ii) Another limitation is the current binding simulation framework Absolut! utilises 3D lattice representation that is based on pre-specified inter-AA distances and 90-degree angles. Such a representation is highly restrictive in many configurations where antibodies can bind to an antigen of interest. We wish to address this in future work, building on a more realistic framework combining docking such as FoldX~\citep{schymkowitz2005foldx} with structure prediction tools like AbodyBuilder~\citep{leem2016abodybuilder}, AlphaFold Multimer~\citep{Evans2021.10.04.463034}. iii) In the current work, we only design the CDRH3 region, which is identified as the most variable chain for an antibody and ignore the folding of other CDR loops that can affect the binding specificity. The above-discussed limitations are promising research questions to extend \texttt{AntBO} that we wish to study in future work.
}
\section{Declaration of interests}
This manuscript is an open-source research contribution by Huawei, Tech R\&D (UK). We release all used resources on GitHub. During the project, Asif Khan held a research intern position, and Alexander-Cowen Rivers held a research scientist position at Huawei. Some authors (Antoine Grosnit, Derrick-Goh-Xin Deik, Rasul Tutunov, Jun Wang and Haitham Bou-Ammar) are currently affiliated with Huawei. Victor Greiff holds advisory board positions in aiNET GmbH and Enpicorm B.V. and is also a consultant for Roche/Genetech.

\section{Code and data availability}
The code of our software \texttt{AntBO} and other used resources are open source on \url{https://github.com/huawei-noah/HEBO/tree/master/AntBO}
\section{Visualisation tools}
We use open source python package matplotlib \citep{hunter2007matplotlib} and seaborn library built on top of matplotlib for the purpose of data visualisation in the paper. For the visualisation in Figure~\ref{fig:AntDesignIntro} we use an open source online tool \url{draw.io}. 
\section{Acknowledgements}
We acknowledge the generous support from the CSTT grant, Huawei's Noah's Ark Lab \& Huawei Tech R\&D (UK), enabling us to conduct this research. The work was carried out and devised on the computing infrastructure provided by Huawei. The collaboration with GrieffLab tightened the numerical analysis and enabled broader interpretations while suggesting additional directions and results. We would also like to thank Simon Mathis, Arian Jamasb and Ryan-Rhys Griffiths from the University of Cambridge for their involvement in the discussion and feedback on the paper.
\bibliography{references}

\onecolumn
\newpage
\section{Methods}
\label{methods_implementation}
\subsection{Introduction to BO and GP}
\paragraph{Gaussian Processes}
\textit{A GP is defined as a collection of random variables, where a joint distribution of any finite number of variables is a Gaussian~\citep{rasmussen2003gaussian}}.\\
Let $f:\gX\rightarrow \sR$ be a continuous function, then the distribution over function $f$ is specified using a GP, that is, $f(\xBRV)\sim GP (\muBRV(\xBRV), \KBRV(\xBRV, \xBRV\prime))$, where $\muBRV(\xBRV)= \mathbb{E} \left[ f (\xBRV) \right] $ is a mean function and $\KBRV(\xBRV, \xBRV^\prime) =\mathbb{E} \left[ (f (\xBRV) - \muBRV(\xBRV) ) (f (\xBRV^\prime) - \muBRV(\xBRV^\prime))\right]$ is a covariance matrix. The standard choice for a mean function is a constant zero $\muBRV(\xBRV)=0$~\citep{rasmussen2003gaussian}, and the entries of a covariance matrix are specified using a kernel function.
% The zero mean doesn’t restrict GP as by just normalising the data, we can completely define the process by its covariance matrix.
By definition, kernel function $\kBRV:\gX\times\gX\rightarrow\sR$ maps a pair of input to a real-valued output that measures the correlation between a pair based on the closeness of points in the input space. As $\gX$ is combinatorial, we need particular kernels to get a measure of correlation, which we introduce in section~\ref{sec:kernel}. 
%For the details of GP modelling we refer readers to Supplementary~\ref{app:gpfit}.\\
\paragraph{GP prediction}
\label{app:gpfit}
Consider $\XBRV = {(\xBRV_i,\yBRV_i)}_{i=1}^{N}$ be a set of training data points and $\XBRV^{\ast} = {(\xBRV_i^{\ast},\yBRV_i^{\ast})}_{i=1}^{N}$ be a set of test data points. To fit a GP, we parameterise kernel hyperparameters and maximise the marginal log-likelihood (MLL) using the data. Specifically, we define $\KBRV(\XBRV,\XBRV)$ as a covariance matrix of training samples, $\KBRV(\XBRV,\XBRV^{\ast})$ and $\KBRV(\XBRV^{\ast},\XBRV)$ are covariance matrix of train-test pairs and vice versa, and $\KBRV(\XBRV^{\ast},\XBRV^{\ast})$ is a covariance matrix of test samples. The final posterior distribution over test samples is obtained by conditioning on the train and test observation as,
\begin{align}
    \mathbf{f}^{\ast}| \XBRV^{\ast}, \XBRV, &\mathbf{y} \sim \gN (\KBRV(\XBRV^{\ast},\XBRV) \KBRV(\XBRV,\XBRV)^{-1} \mathbf{y},\nonumber\\ &\KBRV(\XBRV^{\ast},\XBRV^{\ast}) - \KBRV(\XBRV^{\ast},\XBRV) \KBRV(\XBRV,\XBRV)^{-1} \KBRV(\XBRV,\XBRV^{\ast}))\nonumber
\end{align}
\paragraph{GP Training} We fit the GP by optimising the negative MLL using Adam~\citep{kingma2014adam}. The kernel functions in GPs come with hyperparameters that are useful to adjust the fit of a GP; for example, in a SE kernel described above, we have a lengthscale hyperparameter that acts as a filter to tune the contribution of various frequency components in data. In a standard setup, the optimum value of the hyperparameter is obtained by minimising the negative marginal loglikelihood,
\begin{align}
    - \log p (\yBRV|\XBRV, \theta) &= 0.5 \log |(\KBRV_{\theta}(\XBRV,\XBRV) + \sigma^2 \IBRV)| \nonumber\\
     + 0.5 \yBRV^T (\KBRV_{\theta} &+ \sigma^2\IBRV)^{-1}\KBRV_{\theta}(\XBRV,\XBRV) + 0.5 N \log (2\pi) \nonumber
\end{align} 
where $\theta$ is the set of kernel hyperparameters and $|.|$ is the determinant operator. 

\subsubsection{Kernels}
\label{kernels}
\paragraph{Transformed Overlap Kernel (TK)}
TK defines measure of similarity as $k(\xBRV, \xBRV^\prime) = \exp (\frac{1}{L}\sum_{i=1}^{L} \theta_i \delta(x_i, x_i^\prime))$ where $\{\theta_i\}_{i=1}^{L}$ are the lengthscale parameters that learn the sensitivity of input dimensions allowing GP to learn complex functions.
\paragraph{ProteinBERT Kernel (ProtBERT)}
We utilise a deep kernel for protein design based on the success of transformer architecture BERT. The ProteinBERT~\citep{brandes2021proteinbert} model is a transformer neural network trained on millions of protein sequences over 1000s of GPUs. Such large-scale training facilitates learning of the representation space that is expressive of the higher-order evolutionary information encoded in protein sequences.  We use the encoder of the pre-trained ProteinBERT model followed by a standard RBF kernel to measure the similarity between a pair of inputs.
\paragraph{Fast String Kernel (SSK)~\cite{leslie2004fast}}
\label{app:SSK}
Let $\Sigma^l$ be a set of all possible ordered substrings of length $l$ in the alphabet, $\xBRV$ and $\xBRV^\prime$ be a pair of antibody sequences, then the correlation between the pair is measured using a kernel $k_\theta(.,.)$ is defined as, 
\begin{align}
k_{\theta}(\xBRV, \xBRV)&=\sum_{\yBRV\in\Sigma^l}\phi_{\yBRV}^\theta(\xBRV) \phi_{\yBRV}^\theta(\xBRV^\prime), \nonumber\\
\phi_{\yBRV}^\theta(\xBRV^\prime)=&\theta^{|\yBRV|}_m\sum_{1\leq i_1<,\cdots,\leq i_{k}\leq |\xBRV|} \theta_g^{i_{|\yBRV|} - i_1 + 1} \sI_{\yBRV} [(\xRV_{i_1}^\prime,\cdots,\xRV_{i_{|\yBRV|}}^\prime)]\nonumber
\end{align}
where $\xRV_{i_j}^\prime$ is a length $j$ subsequence of sequence $\xBRV^\prime$, $\theta={\theta_m,\theta_g}$ are kernel hyperparameters, $\theta_m,\theta_g\in[0,1]$ control the relative weighting of long and non-contiguous subsequences, $\sI_{\yBRV} [\xBRV]$ is an indicator function set to $1$ if strings $\xBRV$ and $\yBRV$ match otherwise $0$, and $\phi_{\yBRV}^\theta(\xBRV)$ measures the contribution of subsequence $\yBRV$ to sequence $\xBRV$.
\subsubsection{Acquisition function}
BO relies on the criterion referred to as \textit{acquisition function} to draw new samples (in our problem protein sequences) from the posterior of GP that improve the output of the black-box (binding energy). The most commonly used acquisition function is expected improvement (EI)~\citep{movckus1975bayesian}. EI aims to search for a data point that provides expected improvement over already observed data points. Suppose we have observed $N$ data points $\gD_n = \{(\xBRV_1, f(\xBRV_1)),\cdots,(\xBRV_n,f(\xBRV_n))\}$ then the EI is defined as an expectation over $\gD_n$ under the GP posterior distribution as $\alpha_{\text{EI}} (\xBRV) = \mathbb{E}_{(.|\gD_n)} \left[ \min(\mathbf{f}(\xBRV) - \mathbf{f}(\xBRV^\ast), 0)\right]$, where $\xBRV^\ast = \arg\min_{\xBRV\in \gD_{n}} f(\xBRV)$. There are several other choices of acquisitions we refer readers to~\citep{snoek2012practical, garnett_bayesoptbook_2022, grosnit2021we}.
\begin{algorithm}[t!]
\label{alg:constrainedBO}
\SetAlgoLined
\DontPrintSemicolon
\caption{Antibody Bayesian Optimisation (\texttt{AntBO})}
\KwIn{Objective function $f:X\rightarrow{\mathds{R}}$, number of evaluations $N$, alphabet size of categorical variable $K$.
}
Randomly sample an initial data set $\gD_{1}= {(\xBRV_i, f(\xBRV_i))}_{i=1}^{M}$\\
\For{$i=1,...,N$ }{ 
Fit a GP surrogate $g$ on $\gD_{i}$\\
Construct a $\texttt{CDRH3-TR}_{L_i}(\xBRV^{\ast})$ around the best point $\xBRV^{\ast}=\arg\min_{\xBRV\in\gD_{i}} g(\xBRV)$ using Equation~\ref{eq:trustregion} in the main document.\\
Optimise constrained acquisition,\\
$\xBRV_{i+1}= \arg\min_{\xBRV\in TR(\xBRV^{\ast})} \alpha (\xBRV|\gD_{i})$ \\
Evaluate the black-box $f(\xBRV_{i+1})$ \\
Update the data set $\gD_{i+1} = \gD_{i} \cup (\xBRV_{i+1}, f(\xBRV_{i+1}))$\\
    }
\KwOut{The optimum sequence $\xBRV^{\ast}$}   
\end{algorithm}
\subsection{Implementation details}
We use Python for the implementation of our framework. We run all our experiments on a Linux server with $87$ cores and $12$ GB of GPU memory. We have outlined the hyperparameter used for all the methods in Table~\ref{tab:hyperparam}. For BERT we use a pre-trained ``prot\textunderscore bert\textunderscore bfd" model available from~\cite{brandes2021proteinbert}. We package \texttt{AntBO} as software that comes with an easy interface to introduce a new optimisation algorithm and a black box oracle function. Thus, it offers a platform to investigate new ideas and benchmark them quickly across other methods. We next provide the details of the software. 
\subsection{Software}
\label{subsec:software}
\begin{figure}[t!]
\centering
\subfigure[Architecture of end-to-end framework for black-box optimisation. The architecture divides into four layers. The bottom layer consists of model parameters and experiment configurations which could be defined by the developers. The application layer, pre-written or written by developers, sets up the components for the execution layer. The detail of the execution layer is shown on the right side. The summarise layer collects data every iteration and produces real-time visualisation of the results.]{\includegraphics[width=0.45\linewidth, height=4.75cm]{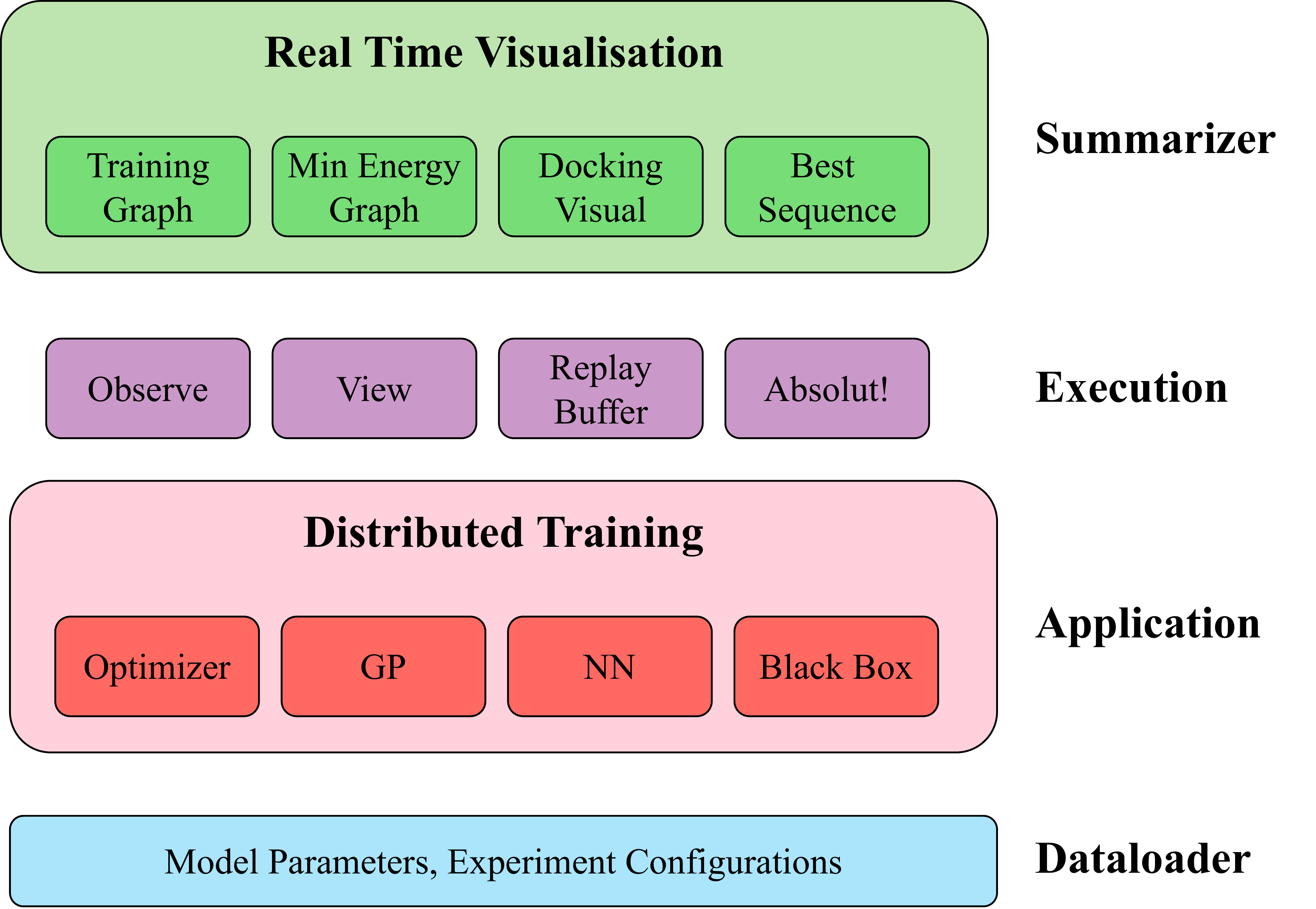}}
\quad
\subfigure[Abstraction within the execution layer. The agent suggests the CDRH3 sequences and passes them into the gym environment. The gym environment evaluates the corresponding binding energies with Absolut!. The agent observes the results and calls the summarisation function to update real-time data. The results are stored in the replay buffer, which can be used to train deep reinforcement learning models. Within the observe function, the model-based agent also optimises the model. The agent then suggests the new CDRH3 sequences in the next iteration.]{\includegraphics[width=0.45\linewidth]{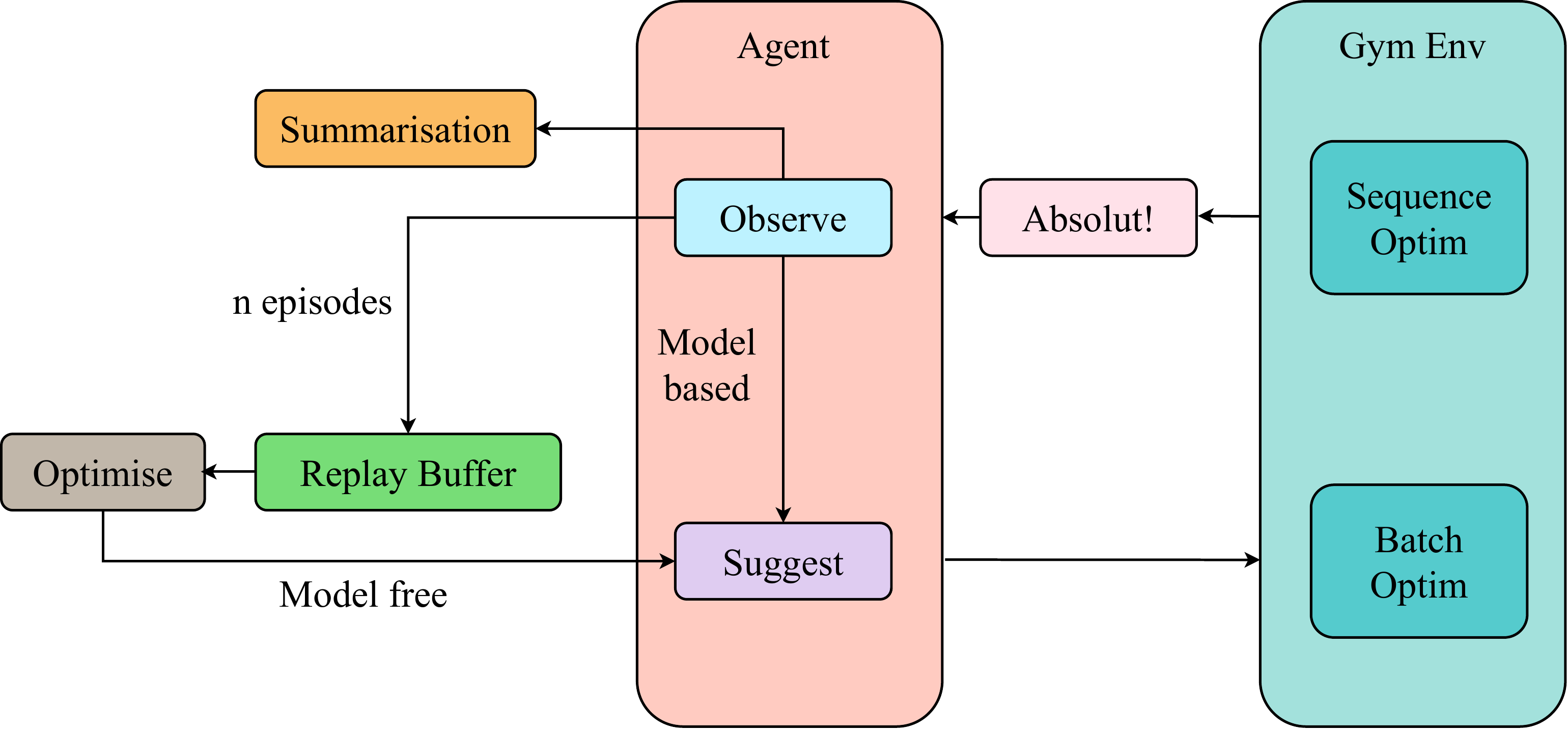}}
\caption{Layout of AntBO packaged as a software. On the left is the architecture of the framework. On the right is the illustration of the execution layer.}
\label{fig:software}
\vspace{-0.5cm}
\end{figure}
\label{subsec:software}
The framework's architecture can be seen in part (a) of Figure~\ref{fig:software}. The dataloader, execution, and summarise layer are abstracted and integrated with the training, leaving only the optimizer for developers to design. The developers could also optionally include Gaussian Process, Neural Network or an arbitrary model to use with the optimiser. The platform has three important features that facilitate training,
\begin{itemize}
  \item Distributed training: Multiple CPU processes for data sampling in a parallel environment, especially useful in low data efficiency algorithms such as deep reinforcement learning. Multiple CPU processes are also utilised to evaluate the binding energy with Absolut, which speeds up the evaluation time. Multiple GPU training for an algorithm that supports the neural network.
  \item Real-time visualisation: Update the training results of the optimizer in real-time. Our framework offers visualisation of the training graph, minimum binding energy obtained so far per iteration, and the corresponding sequence for the minimum binding energy and antigen docking visualisation.
  \item Gym environment: Our framework offers a highly reusable gym environment containing the objective function evaluator via \texttt{Absolut!}. Developers could set the antigen to evaluate and CDRH3 sequences to bind, and the environment returns the binding energy of the corresponding CDRH3 sequences. The gym environment has two options, SequenceOptim and BatchOptim. For SequenceOptim, the agent fills a character in each step until all characters for the CDRH-3 are filled, when the episode stops. For each step, the reward is zero until the last step of the episode, when the CDRH-3 sequences will be evaluated, and the negative binding energy is returned as a reward. The binding energy is negative; hence lower negative binding energy represents a higher reward. For BatchOptim, each episode only has one step, in which the agent inputs the list of CDRH3 sequences of the antigens into the environment and the reward returns are a list of binding energy corresponding to the CDRH-3 sequences. SequenceOptim is useful for seq2seq optimisation, and BatchOptim is useful for combinatorial optimisation.
\end{itemize}
\begin{table}[t!]
    \centering
    \scalebox{0.9}{\begin{tabular}{c|cc}
    \toprule
    \bfseries Algorithm & \bfseries Hyperparameter &\bfseries Value \\
    \midrule
    \begin{tabular}{c} AntBO TK / \\ {AntBO NT} / \\ AntBO SSK / \\     AntBO BERT\end{tabular}  &  \begin{tabular}{c}Acquisition function \\ Nb. of initial points \\Normalise \\ \begin{tabular}{c} Kernel Type TK / \\ {Kernel Type NT} / \\ Kernel Type SSK / \\ Kernel Type BERT \  \end{tabular} \\ Noise Variance \\ Search Strategy \\ 
        {Use trust region TK / NT / SSK / BERT} \\ 
        Trust region Length Min $d_{\min}$ \\ Trust region Length Max $d_{\max}$\end{tabular} & \begin{tabular}{c} Expected Improvement \\ 20 \\ True \\ 
        \begin{tabular}{c}  Transformed Overlap Kernel / \\
         {Transformed Overlap Kernel} / \\
         Fast String Kernel /\\ RBF Kernel with lengthscale on BERT features \end{tabular} \\ 1e-6 \\ CDRH3 {(}trust-region{)} Local Search \\ {Yes / No / Yes / Yes} \\ 1 \\ 30 \end{tabular} \\
    \hline
        COMBO & \begin{tabular}{c} Batch size \\ Nb. of initial points \\ GP-parameters slice sampling steps \\
       Acquisition function \\ Nb. of random samples for BFLS \\ Nb. of initial points for BFLS  \end{tabular}  & \begin{tabular}{c} 1 \\ 20 \\ 100 (init) / 10 (refine) \\ Expected Improvement \\ 20,000 \\ 20  \end{tabular} \\
    \hline
        HEBO &  \begin{tabular}{c} Batch size \\  Surrogate Model \\Acquisition Class\\ Acquisition Optimiser\\ Population Size \\ Optimiser Nb. of Iterations \\ Optimiser ES \end{tabular}  &  \begin{tabular}{c} 1 \\ Gaussian Process \\ Evolution Optimiser \\MACE\\ 100 \\ 100 \\ NSGA-II \end{tabular} \\
    \hline
        TURBO & \begin{tabular}{c} Batch size \\ Nb. of trust regions \\  Trust region Length Min \\ Trust region Length Max \\ Trust region Length Init \\ $\tau_\text{succ}$ \\ $\tau_\text{fail}$ \\  Max Cholesky Size  \\ GP fit - Optimiser \\ GP fit - Training Steps \\ GP fit - Learning Rate \\ Nb. of Thompson Samples\end{tabular} & \begin{tabular}{c} 1 \\ 1 \\  $2^{-7}$ \\ 1.6 \\ 0.8\\ 3 \\ $d$ \\ 2000 \\ Adam \\ 50 \\ 0.1 \\ $\min(100 d, 5000)$ \end{tabular} \\
    \hline
        GA & \begin{tabular}{c} Population size \\ Nb. of iterations \\ Nb. of parents \\Nb. of elite \\ Crossover type \\ Crossover probability \\ Elite ratio \\ Mutation probability \end{tabular} & \begin{tabular}{c} 40 \\ 5 \\ 16 \\ 6 \\ uniform \\ 1. \\ 0.15 \\ $1/d$   \end{tabular} \\
    \hline
        RS & \begin{tabular}{c}  Nb. of iterations \\ Sampling type \end{tabular} & \begin{tabular}{c}  200 \\ uniform \end{tabular} \\
    \hline
        {LamBO} & 
       \begin{tabular}{c}  {Query batch size ($b$)} \\ {Batch set size ($|\mathcal{X}_\text{base}|$)} \\ {Nb. of initial points ($|\mathcal{D}_0|$)} \\ {Surrogate model} \\ {Acquisition function)} \\ {Encoder} \end{tabular} &  \begin{tabular}{c}  {$1$} \\ {$16$} \\ {$200$} \\ {Exact GP (\texttt{single\_task\_exact\_gp})} \\ {Expected Improvement} \\ {\texttt{mlm\_cnn}} \end{tabular} \\
    \bottomrule
    \end{tabular}}
    \caption{Hyperparameter Configuration of different optimisation methods.}
    \label{tab:hyperparam}
\end{table}
\subsection{Baseline methods}
\label{subsec:baseline}
In this section we discuss details of all the baseline methods we use for comparison. 
\subsubsection{Random search}
Given a computational budget of $N$ black-box function evaluations in a constrained optimisation setting, RS samples $N$ candidates that satisfy the specified constraints and evaluates the black-box function at those samples. A best candidate is the one with the minimum cost.
\subsubsection{Baseline BO methods}
\paragraph{HEBO}
The Heteroscedastic and Evolutionary Bayesian Optimisation solver  (HEBO~\cite{cowen2020empirical}) is the winning solution of the NeurIPS 2020 black-box optimisation (BBO) challenge \cite{turner2021bbox}. HEBO is designed to tackle BBO problems with continuous or categorical variables, dealing with categorical values by transforming them into one-hot encodings. Efforts are made on the modeling side to correct the potential heteroscedasticity and non-stationarity of the objective function, which can be hard to capture with a vanilla GP. To improve the modeling capacity, parametrised non-linear input and output transformations are combined to a GP with a constant mean and a Mat\'ern-3/2 kernel. When fitting the dataset of observations, the parameters of the transformations and of the GP are learned together by minimising the negative marginal likelihood using Limited-memory BFGS (LBFGS) optimiser.
When it comes to the suggestion of a new point, HEBO accounts for the imperfect fit of the model, and for the potential bias induced by the choice of a specific acquisition function, by using a multi-objective acquisitions framework, looking for a Pareto-front solution. Non-dominated sorting genetic algorithm II (NSGA-II), an evolutionary method that naturally handles constrained discrete optimisation, is run to jointly optimise the \textit{Expected Improvement}, the \textit{Probability of Improvement}, and the \textit{Upper Confidence Bound}. The final suggestion is queried from the Pareto front of the valid solutions found by NSGA-II that is run with a population of $100$ candidate points for $100$ optimisation steps.\\
HEBO results presented in this paper are obtained by running the official implementation by \cite{cowen2020empirical} at \url{https://github.com/huawei-noah/HEBO/tree/master/HEBO}.
\paragraph{TuRBO}
To tackle optimisation of high-dimensional black-box functions, BO solvers face the difficulty of finding good hyperparameters to fit a global GP over the entire domain, as well as the challenge of directly exploring an exponentially growing search space. \cite{eriksson2019scalable} introduces the use of local BO solvers to alleviate the above issues. The key idea is to use local BO solvers in separate subregions of the search space, leading to a trust region BO algorithm (TuRBO). A TR is a hyperrectangle characterised by a centre point and a side length $L$ similar to what we describe in Section~\ref{sec:cdrh3trustcombo}. A local GP with constant mean and Mat\'ern-5/2 ARD kernel fits the points lying in the TR better to capture the objective function's behaviour in this subdomain. The GP fit is obtained by optimising the negative MLL using Adam~\cite{kingma2014adam}. The size of the TR is adjusted dynamically as new points are observed. The side length $L$ is doubled (up to $L_\text{max}$) after $\tau_\text{succ}$ consecutive improvements of the observed black-box values, and is halved after $\tau_{fail}$ consecutive failures to find a better point in the TR. The TR is terminated whenever $L$ shrinks to an $L_\text{min}$ value, and a new TR is initialised with a side size of $L_\text{init}$. The next point to evaluate is selected using the Thompson Sampling strategy, which ideally consists of drawing a function $f$ from the GP posterior and finding its minimiser. However, it is impossible to draw a function directly over the entire TR; therefore, a set of $\min(100 d, 5000)$ candidate points covering the TR is used instead. Function values are sampled from the surrogate model's joint posterior at these candidate points. The candidate point achieving the lowest sample value is acquired. Our experiments only acquire suggested points that fulfil the developability constraints.

In our experiments, we rely on the TuRBO implementation provided in the BBO challenge \cite{turner2021bbox} codebase at \url{https://github.com/rdturnermtl/bbo_challenge_starter_kit/tree/master/example_submissions/turbo}.
\begin{algorithm}[t!]
\label{alg:ga}
\SetAlgoLined
\DontPrintSemicolon
\caption{Genetic Algorithm}
\KwIn{Black box function $f:X\rightarrow{\mathds{R}}$, Constraint function $C:X\rightarrow {0, 1}$, Maximum number of iterations $N_{iter}$, Population size $N_{pop}$, Number of elite samples $N_{elite}$, Crossover probability $p_{c}$, Mutation probability $p_m$}
\KwOut{Best performing sample}

\nl $P^{0} =rejectionSampling(C)$ \tcp*{Sample initial population}
\nl $F^{0} \leftarrow f(P^{0})$ \tcp*{Evaluate initial population}
\For{$i=1,...,N_{iter}$ }{ 
    \nl $P^{i+1} = []$ \tcp*{Initialise next population with an empty list}
    \nl $Q^{i} = []$ \tcp*{initialise list of parents}
    \For{$j=1,...,N_{elite}$}{
        \nl $p^j \leftarrow$ sample with $j^{th}$ highest fitness from $P^{i}$ \tcp*{Get sample with the next highest fitness}
        \nl $P^{i+1} \leftarrow P^{i+1} \cup p^j$ \tcp*{Add this sample to the next population}
        \nl $Q^{i} \leftarrow Q^{i} \cup p^j$ \tcp*{Add this sample to the list of parents}
        }
    % Not sure on correct way to say j increases in steps of 2
    \For{$j=N_{elite} +1, ..., N_{pop}$ where $j$ increases in steps of 2}{
        \nl $q_{1}, q_{2} \sim Q^{i}$ \tcp*{Randomly sample two parents}
        \nl $constraint\_satisfied = False$ \;
        \While{not $constraint\_satisfied$}{
            \nl $\eta_{1}, \eta_{2} = crossover(q_{1}, q_{2}, p_{c})$ \tcp*{Perform crossover to generate two offsprings}
            \nl $\eta_{1}, \eta_{2} \leftarrow mutate(\eta_{1}, p_{m}), mutate(\eta_{2}, p_{m})$ \tcp*{Mutate both offsprings}
            \nl $constraint\_satisfied = C(\eta_{1}) \wedge C(\eta_{2})$ \tcp*{Check that both offsprings satisfy all constraints}
            }
        \nl $P^{i+1} \leftarrow P^{i+1} \cup \eta_{1} \cup \eta_{2}$ \tcp*{Add offsprings to new population}
        }
    \nl $F^{i+1} \leftarrow f(P^{i+1})$ \tcp*{Evaluate new population}
    }
\end{algorithm}
\paragraph{COMBO}
To adapt the BO framework for combinatorial problems, \cite{Oh_2019} proposed to represent each element of the discrete search space as a node in a combinatorial graph. Then a GP surrogate model is trained for the task of node regression using a diffusion kernel over the combinatorial graph. However, the graph grows exponentially with the number of variables, making it impractical to compute its diffusion kernel. To address this issue, the authors express the graph as a cartesian product of subgraphs.
This decomposition allows the computation of a graph diffusion kernel as a cartesian product of kernels on subgraphs. The efficient computation of diffusion kernel is done using Fourier transform. The hyperparameters of the GP model, such as kernel scaling factors, signal variance, noise variance, and constant mean value, are obtained using 100 slice sampling steps at the beginning and ten slice sampling iterations afterwards.
Once we obtain the GP fit, it remains to optimise an acquisition function over the combinatorial space, which is done by applying a breadth-first local search (BFLS) from $20$ starting points selected from $20,000$ evaluated random vertices. Since COMBO does not support constraints on the validity of the suggested sequences, we modify the acquisition optimisation to incorporate the \texttt{CDRH3-TR} introduced in~\ref{eq:trustregion}. We use the default hyperparameters that we provide on Table~\ref{tab:hyperparam} and add constraints handling to the official implementation by \cite{Oh_2019} at \url{https://github.com/QUVA-Lab/COMBO}.

\subsubsection{Genetic algorithm}
Genetic algorithms (GAs) are inspired by Charles Darwin's theory of natural selection. The idea is to use probabilistic criteria to draw new population samples from the current population. This sampling is generally done via crossover and mutation operations \cite{Sastry2005}. Overall the primary operations involved in GA are: encoding schemes, crossover, mutation, and selection, respectively~\cite{reviewGA}. For encoding, we use a general ordinal encoding scheme that assigns a unique integer to each AA—inspired from binary encoding where each gene represents integer 0-1 or hexadecimal that represents integer 0-15 (or 0-9, A-F)~\cite{binaryGA}. Specific to our work, we express each gene by a letter of CDRH3 sequences ranging from (0-19). For selection, we use the elitism mechanism \cite{elitismGA}, which preserves a few best solutions in the current population to the next population. Our mutation operator is inspired by the most commonly used bit flipping mutation \cite{bitGA} that flips a bit of each gene with a given probability. Instead, we randomly replace a gene from 0-19 as our range is different. Finally, for crossover, we use a uniform crossover, which suggests unbiased exploration and better recombination \cite{reviewGA}. The pseudocode of a GA is illustrated in Algorithm~\ref{alg:ga}.

\clearpage

\newpage
\appendix
\section*{Supplementary Information}

\section{\texttt{Absolut!} A binding affinity computation framework}
\label{subsec:binding}
\texttt{Absolut!}~\citep{robert2021one} is a state-of-the-art \textit{in silico} simulation suite that considers biophysical properties of antigen and antibody to create a simulation of feasible bindings of antigen and antibody. Although \texttt{Absolut!} is not able to directly generate antibody-antigen bindings at the atomic resolution, and therefore to predict antibody candidates directly. However, using \texttt{Absolut!}, we can develop methods in the simulation world and later employ the best method in the complex real-world scenario, with the knowledge that this method already performed well on the levels of complexity already embedded into \texttt{Absolut!} datasets. This feature of \texttt{Absolut!} makes it an ideal black-box candidate for the antibody design problem. However, we note that AntBO is, in principle, agnostic to the choice of the black-box oracle used and can be adapted to other in silico or experimental oracles provided they can compute or determine binding affinity or any other criteria relevant for antibody design. \texttt{Absolut!} performs the computation of binding affinity in three main steps, i) antibody-antigen lattice representation, ii) discretisation of antigen and iii) binding affinity computation. We next introduce the main steps of binding affinity computation in \textbf{\texttt{Absolut!}}.
\subsection{Discretisation of Antigen}
The \texttt{Absolut!} suite utilises Latfit~\citep{Mann-Saunders:12, Mann_CPSPweb_2009} to transform a PDB structure of an antigen into a 3D lattice coordinates position. The PDB structure represents each residue in a protein sequence using 3D coordinates. The Latfit maps these coordinates to a discretised lattice position by optimising dRMSD (Root Mean Square Distance) between the original PDB structure and many possible lattice reconstitutions of the same chains. Specifically, for a sequence of length $L$, Latfit first assigns a lattice position to a starting residue and then enumerates all neighbouring sites to select the one with the best dRMSD to the PDB coordinate of the next residue. The generated nascent lattice structures are rotated to better match the original PDB before adding the next AA. This process is repeated sequentially, and at each step, Latfit keeps track of $N$ best structures of length $K$ to find the best position of the next residue. 
\subsection{Antibody-antigen binding representation}
\texttt{Absolut!} uses the Ymir~\cite{robert2021ymir} framework to represent the protein structures as a 3D lattice model. 
A protein's primary structure is a sequence of amino acids (AA). In a 3D lattice structure, each AA can occupy a single position, and the consecutive AAs occupy the neighbouring sites.
This layout form only permits a fixed inter-AA distance with joint angles of 90 degrees.\\
The structure of the protein is specified with the help of a starting position in the grid and a sequence of relative moves (straight (S), up (U), down (D), left (L), right (R)) that determine the next AA position. The first step is to define a coordinate system with the starting point as an observer and the next move relative to the observer to specify the sequence of moves. There is also a possibility of backwards (B) for the first move that is not allowed for other positions to prevent any collision.
\subsection{Computation of antibody-antigen binding affinity}
In this stage, the lattice structure of two proteins is used to compute their binding affinity. Since the structure of the antibody is not known apriori for a specific antigen, all possible foldings of CDRH3 are generated recursively using the algorithm proposed in~\cite{robert2021ymir} and stored in the memory. As the number of possibilities of folding grows combinatorially with the length of a sequence, \texttt{Absolut!} restricts the size of the CDRH3 sequence to $11$ and limits the search to structures with a realistic minimum of contact points ($10$) to the target antigen.\\
After we obtain the lattice structure of an antigen and the list of pre-computed structures for the CDRH3 sequences, the binding affinity of one structure is described as a summation of three terms, a) \textit{binding energy} the interaction between residues of antibody and residues of antigen, b) \textit{antibody folding energy} the interactions within the residues of antibody, and c)  \textit{antigen folding energy} the interaction within the residues of antigen. Since the antigen structure is fixed apriori, the third term is constant and can be ignored.\\
Consider a pair of lattice positions and residues of an antigen sequence  $(\SBRV, \RBRV)$ and of an antibody sequence $(\GBRV, \KBRV)$. The binding energy $\EBRV_{\text{bind}}$ is defined as a sum of all interaction potential,
\begin{align}
\label{eq:bind}
\EBRV_{\text{bind}} &= \sum_{k=1}^{L_G}\sum_{j=1}^{L} \sI (S_j, K_k) \gA ({R_j}, G_k)
\end{align}
and the folding energy $\EBRV_\text{fold}$ of an antibody is defined as a sum of intra-bonds between its AAs,
\begin{align}
\EBRV_\text{fold} &= \sum_{j=1}^{L}\sum_{k=1}^{L} \sI (K_j, K_k) \gA (K_j, K_k)
\end{align}
where $\gA (., .)$ is an interaction potential of residues determined via Miyazawa-Jernigan interaction potential~\cite{miyazawa1996residue} and  $\sI(a, b) $ is an indicator function that takes the value $1$ if $a$ and $b$ are non-covalent neighbors in the lattice otherwise $0$. For the evaluation of an arbitrary CDRH3 sequence, the precomputed structures are filled one by one with residues of CDRH3, and their total energy is computed as $\EBRV_\text{total} = \EBRV_\text{fold} + \EBRV_\text{bind}$, this step is known as \textit{exhaustive docking}. The best structure is then selected using the minimum total energy criterion. \texttt{Absolut!} does this computation for sequences of length $11$; if the CDRH3 is of size greater than $11$, the same process is repeated for all subsequences of length $11$ with a stride of $1$ from left to right. Altogether, the total energy of an antibody-antigen structure determines its stabililty, and the binding energy is the term that represents the energy score (binding affinity), that aims to be minimized in this work.  

\begin{table}[t!]
  \centering
  \caption{We run multiple trials using a distinct set of initial points in GP. This way, in acquisition maximisation, local search explores a different trajectory on the optimisation landscape, converging to a separate local optimum. We report CDRH3 sequence and a binding energy across trials. The first, column is a PDB id underscore chain of an antigen along with the name of associated disease.}
  \label{tab:antboCDRH3s}
  \begin{adjustbox}{width=\textwidth}
    \begin{tabular}{lllllllllll}
    \toprule
    \diagbox[innerwidth=1.75cm,font=\footnotesize]{PDB}{Trial} &           1 &           2 &           3 &           4 &           5 &           6 &           7 &           8 &           9 &           10 \\
    \midrule
    1ADQ\_A &  LFVFFLLLLEI &  FFFFLLFLLLL &  FFIFFLLFLIL &  LYMRIYLFFLM &  MFFILGMLLFF &  MFVFLRLWFLL &  FLPILFLVLIL &  VFEFIFPCSLV &  MFPCFFFLLLI &  MLLLLLFFLLL \\
    IGG4 Fc Region &&&&&&&&&&\\
    Total Energy       &      -111.05 &      -112.33 &      -112.08 &      -110.52 &      -108.15 &      -109.16 &      -110.38 &      -107.65 &      -109.15 &      -110.58 \\
    1FBI\_X &  WFFFVILNFFF &  FLLKMVFLLLL &  LFFSLLFWLFL &  FFLELFLFFFL &  LFLMFFLDLFF &  LFLWLVLWLFF &  MYFPLIFFMFL &  LFLFIMLPFLL &  IFLTLVLFLFF &  LFIFFVFAFFL \\
     Guinea Fowl Lysozyme &&&&&&&&&&\\
    Total Energy        &      -111.35 &      -112.28 &      -114.86 &      -116.28 &      -114.69 &      -114.23 &      -108.83 &      -114.38 &      -114.14 &      -114.65 \\
    1H0D\_C &  YYLMFRSFMFF &  IFFLGLFLFWL &  YFLFTRLLVLW &  DLLVIRCFWWF &  LGIVWLHLFII &  YLCVHSWLVLV &  LMFNLCKLFYL &  FFWLKFFLLCL &  MIGLFEWISIM &  FRLEILFILLM \\
    Angiogenin &&&&&&&&&&\\
    Total Energy        &       -92.57 &       -90.85 &       -97.34 &       -91.89 &       -86.71 &       -95.96 &       -94.32 &       -90.36 &       -95.44 &       -85.61 \\
    1NSN\_S &  GFKEDLCLLWI &  IFLQLLLFLML &  LFFIRISLFFF &  KIMFILLWCKL &  LMFFDFFFLFE &  LLAFEMDFALL &  DLLIFLDFLFL &  YILFIFSFFLI &  FGLLLLFWVLL &  LFLWLLFFLWL \\
    SNASE: Staphylococcal nuclease complex &&&&&&&&&&\\
    Total Energy        &      -102.07 &      -103.99 &      -106.07 &      -101.73 &      -105.26 &      -110.36 &      -106.28 &      -104.46 &      -101.12 &       -105 \\
    1OB1\_C &  FFWCLKLFGLF &  WTFIHFPVYFM &  LWFLQFFVLVL &  LDFWFLKLFWL &  FDFYFLNFFCL &  LCVTCFELFYI &  WNLIFFSFFVF &  LQLDFYHMFLI &  FWLFFFVIFWF &  LYLLTLLYLDF \\
    MSP1: Merozoite Surface Protein 1 &&&&&&&&&&\\
   Total Energy         &       -98.41 &      -103.51 &      -111.88 &      -111.48 &       -111.1 &      -105.77 &      -108.71 &      -104.07 &      -110.59 &      -111.13 \\
    1WEJ\_F &  FFFLLFRFFVF &  FFFLYMLFFIL &  MHFHLLLLYWL &  CILFFFCFFLL &  LLLFFFYFLIL &  LCFNFLLLEWI &  FHFFECFLLLI &  HIFAHLLEFHL &  FSLKLLLFDFF &  FFIFLFIFLLL \\
    CYC: Cytochrome C &&&&&&&&&&\\
    Total Energy        &       -87.81 &       -86.22 &       -87.81 &        -85 &       -88.23 &       -87.77 &       -86.68 &       -80.61 &       -89.19 &       -87.95 \\
    2YPV\_A &  FFFLLLLFLLL &  LFFMLALLACA &  FMYMFVFLHFS &  FFFFLFFLLLL &  YFVLEFLWFFQ &  GLHMYLLVFLA &  FLFLLLLFKIL &  FLFLLMSFLLL &  LLFAFCLLTLA &  LLLILLLIFLF \\
    fHbp: factor H binding protein &&&&&&&&&&\\
    Total Energy        &       -106.1 &      -113.96 &      -112.03 &      -105.99 &      -101.59 &      -107.53 &      -105.75 &      -105.24 &      -114.44 &      -104.89 \\
    3RAJ\_A &  FFLFLILLMFL &  HFYLFLITIWC &  CFLGVVLIVFF &  LWWLLMLLILL &  LLIFLILMIFF &  IWMIFIIFILL &  CFFHILVLMIW &  MFFLVMFLMLL &  HLLLLLTLLLL &  FFFIILFFLLI \\
    CD38: ADP-Ribosyl cyclase 1 &&&&&&&&&&\\
    Total Energy        &      -124.19 &      -113.21 &       -118.3 &      -121.92 &      -122.18 &       -119.2 &      -115.79 &      -118.86 &      -116.47 &      -123.02 \\
    3VRL\_C &  CLRIVLLFFFL &  FFLFLFLRLFL &  LFFPLFLLIII &  FLPFIFFLLVL &  LVLLLLVFKFF &  CICLIFFFILI &  LIFLLFMFRLL &  LLKFVLLLLIL &  FLLSFFFLMLI &  LFFFLFVMPLI \\
    HIV Gag protein&&&&&&&&&&\\
    Total Energy        &      -115.29 &      -112.36 &      -110.31 &      -116.69 &      -116.18 &      -114.66 &      -117.54 &      -117.49 &      -110.36 &      -115.45 \\
    2DD8\_S &  FWNFRHFYILL &  LCFILLKFDIL &  FFGMNFLYLFL &  WFLWIFDFSLL &  LLKFSFFFLLL &  LFLIILQFEFF &  LLLAGLLTFAL &  LLFHFLQFGFI &  LFYFKFWIFIF &  IFAHLYVYVFL \\
    SARS-CoV Virus Spike glycoprotein &&&&&&&&&&\\
    Total Energy        &      -123.31 &       -129.2 &      -127.68 &       -127 &      -132.27 &      -131.08 &      -123.43 &      -129.03 &      -129.64 &      -118.76 \\
    1S78\_B &  LLLFFLFCFLL &  LLLYLFFLFLF &  FLCIICFLFLM &  LLMIFCIFLFM &  LLLLFLLCFLL &  HWLLIHIFFLL &  LDFLALLLLFL &  FFFWFLMLLFL &  FFLHLLFLLFL &  VILLILWWFLL \\
    HER2: Receptor protein-tyrosine kinase erbB-2 &&&&&&&&&&\\
     Total Energy       &       -143.8 &      -142.38 &      -138.92 &      -139.57 &      -144.02 &      -137.05 &      -140.84 &      -142.19 &      -141.67 &       -138 \\
    2JEL\_P &  WFTLIFLWLDI &  LFFNLLLLIWL &  FLLLLFFGLFF &  LILLLNFFLFL &  FLLLFLLFLFL &  MKIYNLLYLLF &  LFFFLFFRLFI &  FNFFEFFRLLL &  WFLILLTLILF &  LFFLSFLELWF \\
    ptsH: Phosphocarrier protein HPr &&&&&&&&&&\\
    Total Energy        &       -84.18 &       -82.36 &       -83.93 &       -83.72 &       -83.93 &       -82.34 &       -83.88 &       -86.34 &       -81.42 &       -82.95 \\
    \bottomrule
    \end{tabular}
\end{adjustbox}

\end{table}
%and finetune it for $10$ epochs on CDRH3s of $12$ antigens available from \texttt{Absolut! 6.9M} database. 
%We then use this model as a feature warping function that transforms categorical sequences to continuous space. Using a kernel in the latent space of BERT allows us to leverage the prior
\label{apsec:hyper}
\section{Extended results}
\label{apsec:results}
\begin{figure}
\centering
\subfigure[]{\includegraphics[width=0.9\linewidth]{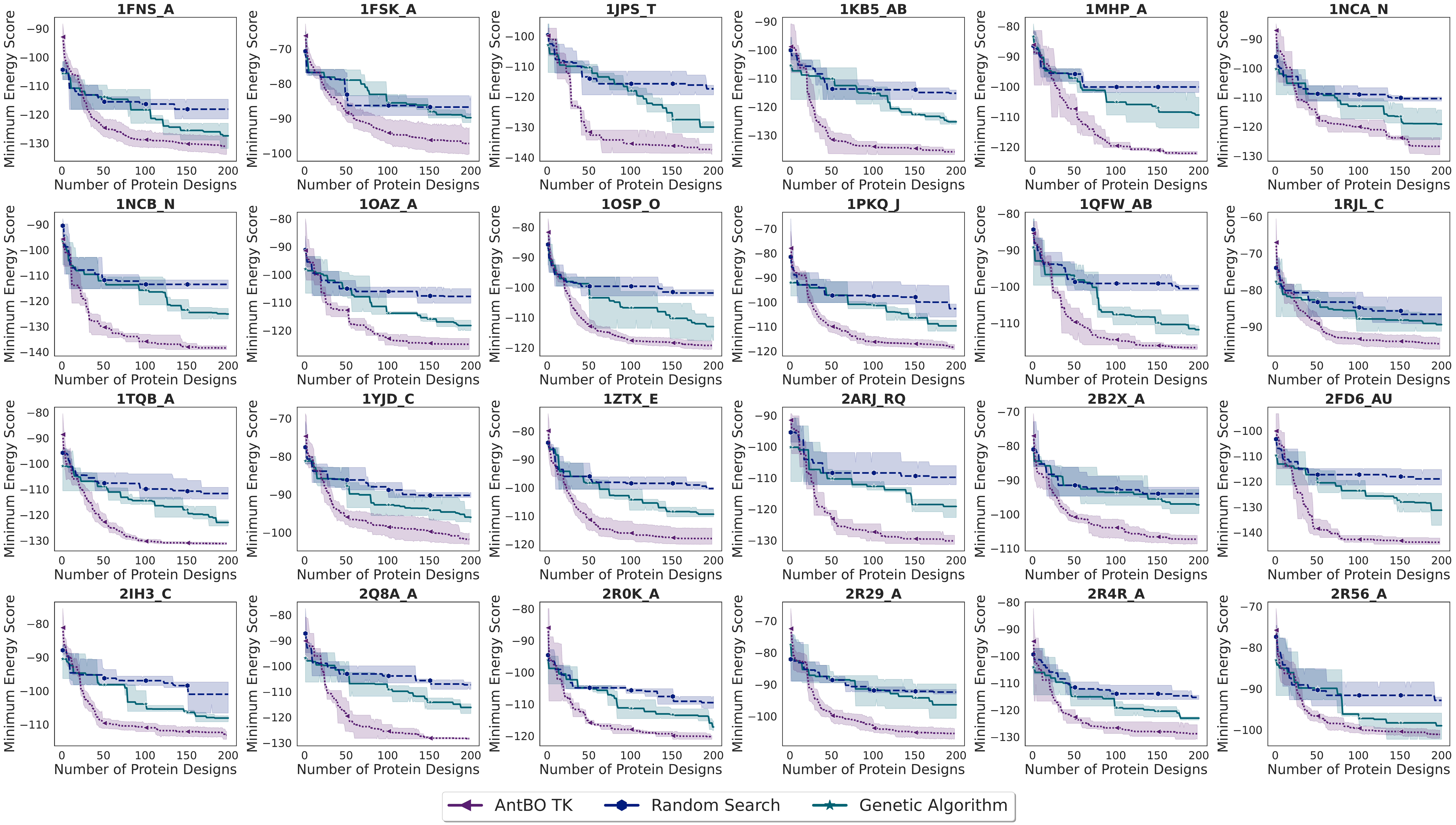}}
\subfigure[]{\includegraphics[width=0.9\linewidth]{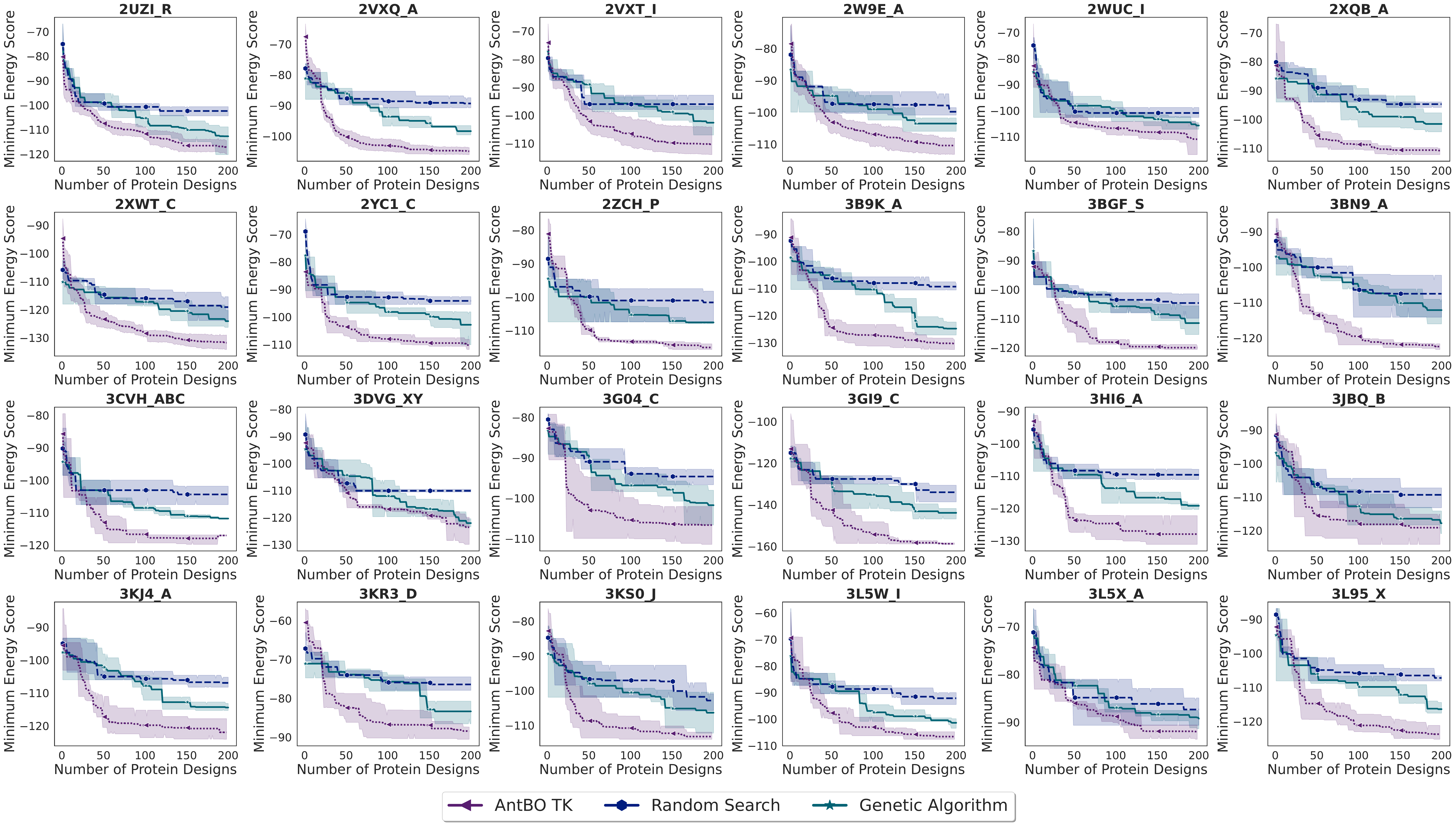}}
\caption{Binding energy vs number of protein designs.}
\label{fig:binding_vs_funct_evals_app_2}
\end{figure}

\begin{figure}
\centering
\subfigure[]{\includegraphics[width=0.9\linewidth]{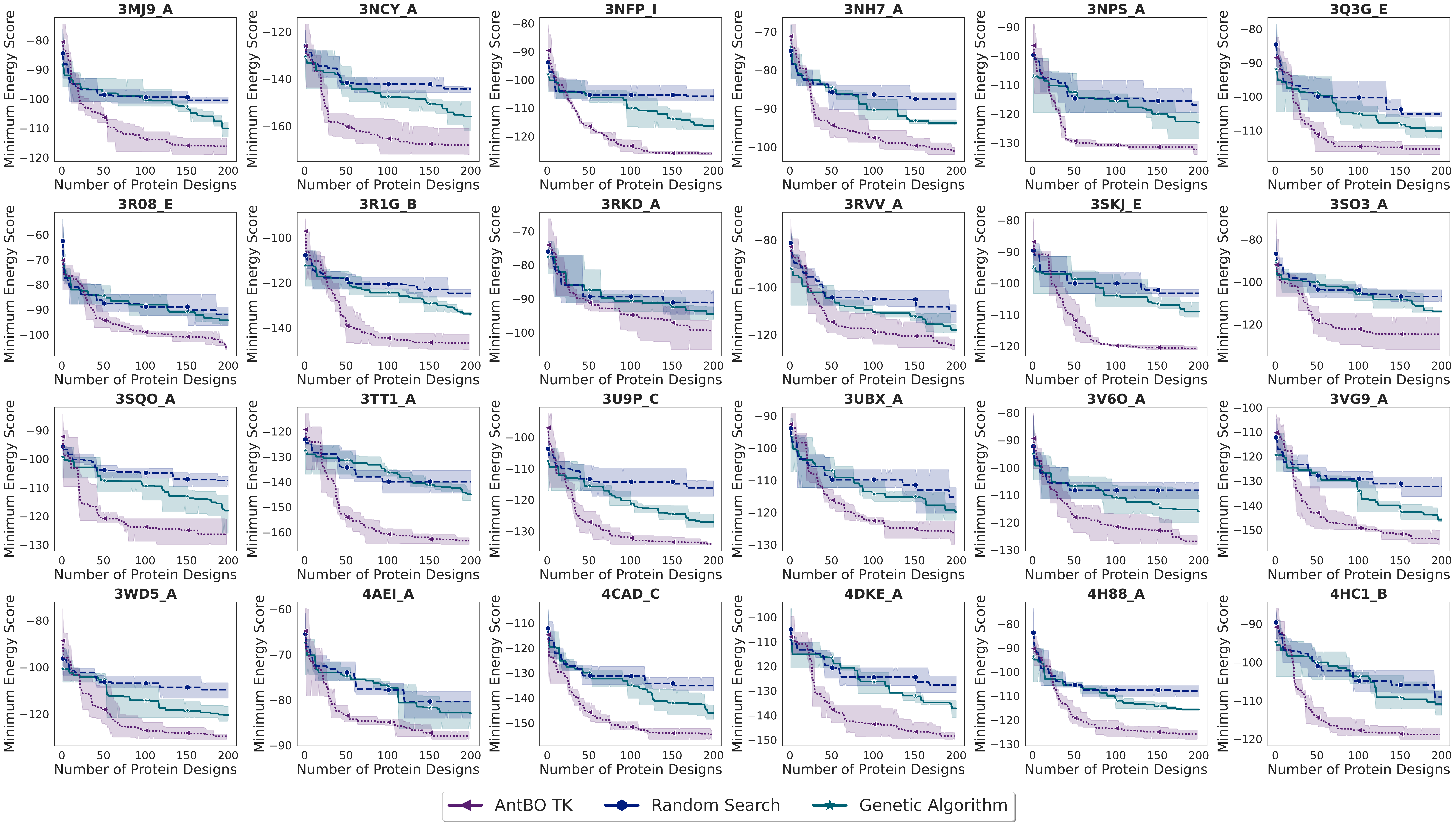}}
\subfigure[]{\includegraphics[width=0.9\linewidth]{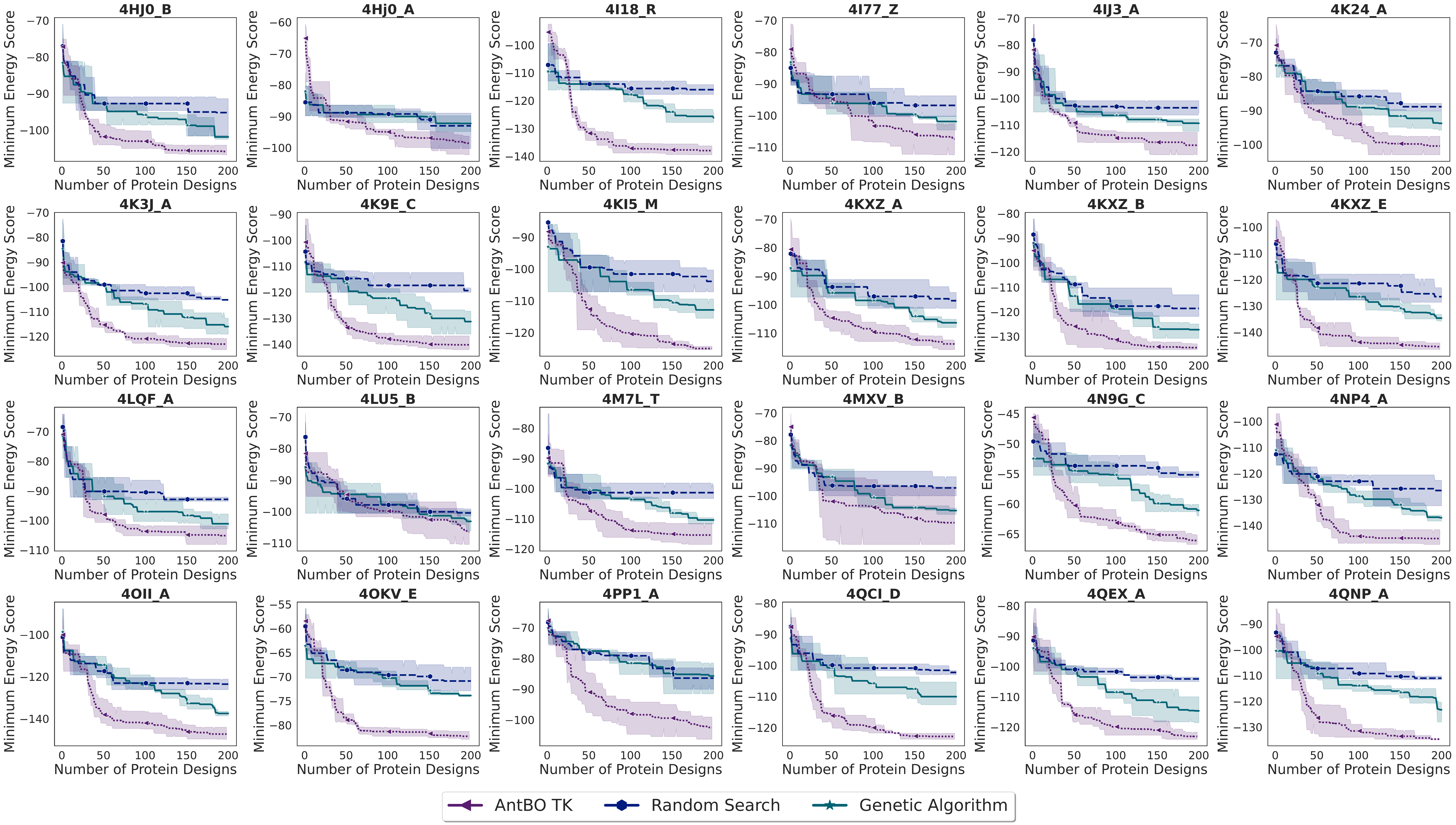}}
\caption{Binding energy vs number of protein designs.}
\label{fig:binding_vs_funct_evals_app_3}
\end{figure}

\begin{figure}
\centering
\subfigure[]{\includegraphics[width=0.9\linewidth]{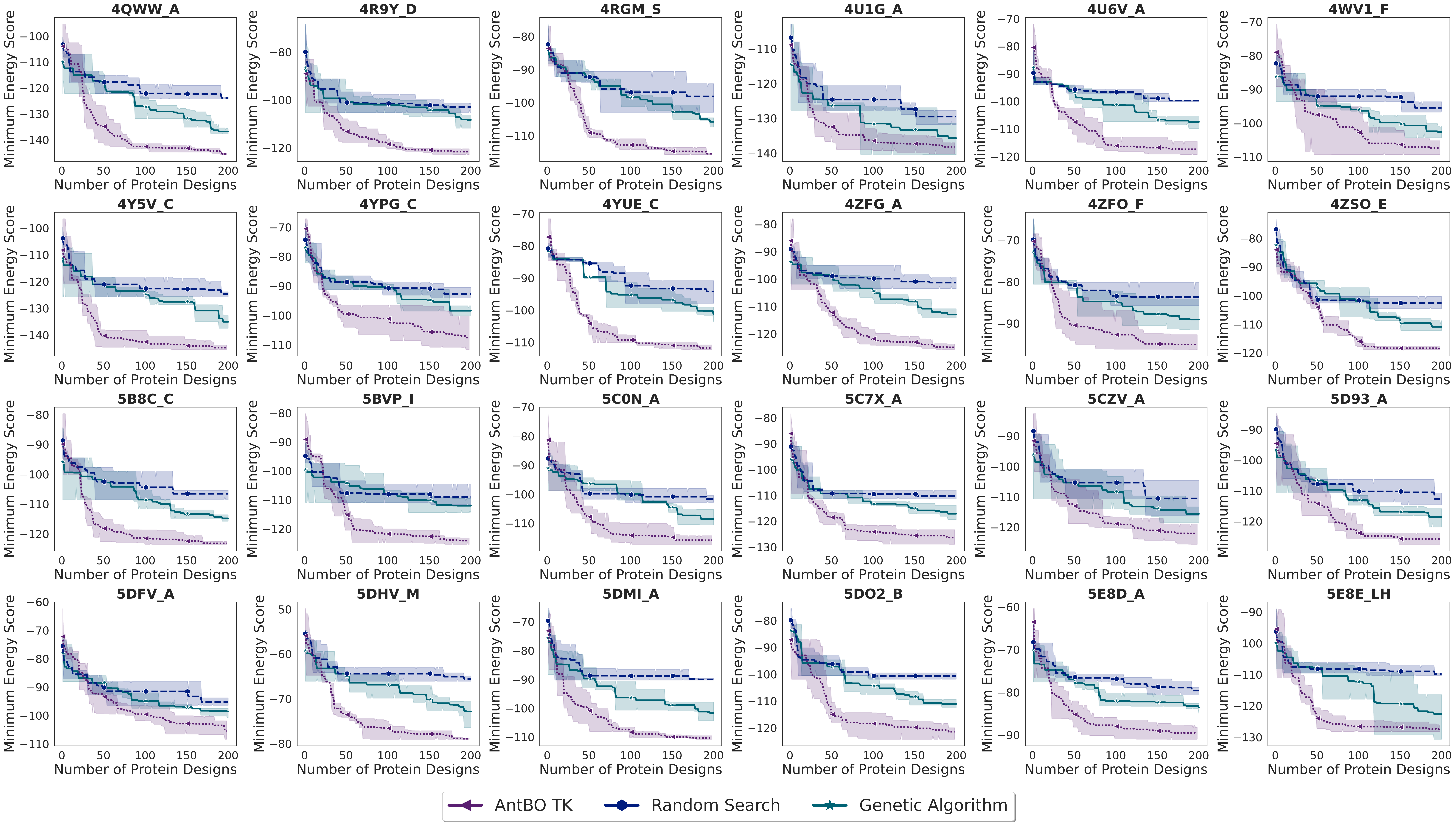}}
\subfigure[]{\includegraphics[width=0.9\linewidth]{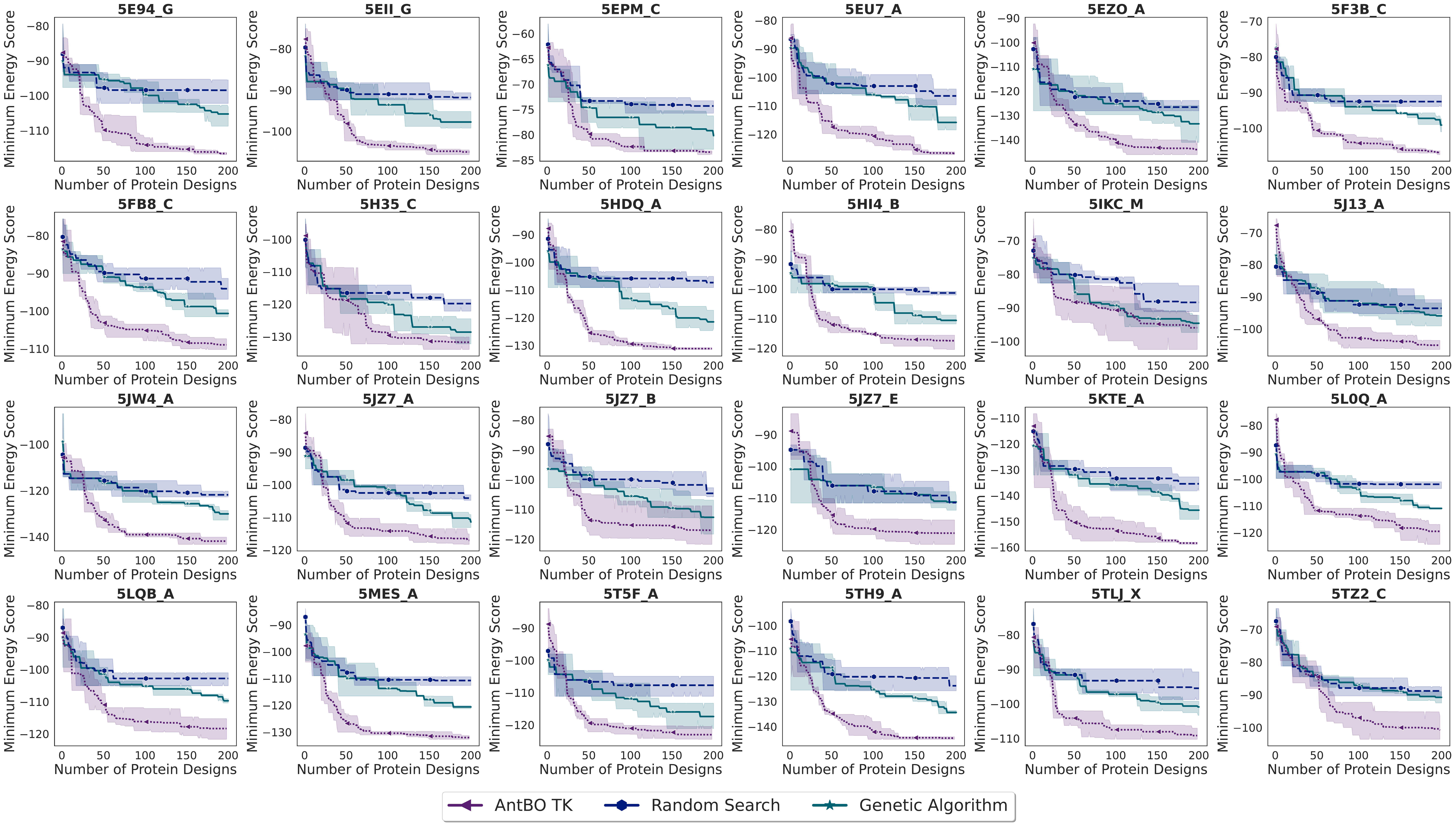}}
\caption{Binding energy vs number of protein designs.}
\label{fig:binding_vs_funct_evals_app_6}
\end{figure}

Here, we provide additional results to demonstrate the performance of AntBO. We first report the convergence curve of a number of proteins evaluated by \texttt{Absolut} vs total energy (or binding affinity). We compare the best performing BO method \texttt{AntBO TK} with other baselines. The results are described in Figure~(\ref{fig:binding_vs_funct_evals_app_2},\ref{fig:binding_vs_funct_evals_app_3},\ref{fig:binding_vs_funct_evals_app_6}). 
We observe \texttt{AntBO} consistently outperforms other baseline methods suggesting our approach can efficiently suggest antibodies for various antigens of interest. On average, we want to emphasise that it only takes approximately $38$ evaluations to find an optimum sequence. This sample efficiency shows \texttt{AntBO} is a vital development toward real-world antibody design.

\begin{figure}[ht!]
    \centering
        \subfigure[1ADQ (A)]{\includegraphics[clip, trim=11cm 3.5cm 9cm 3.5cm, width=0.9\textwidth]{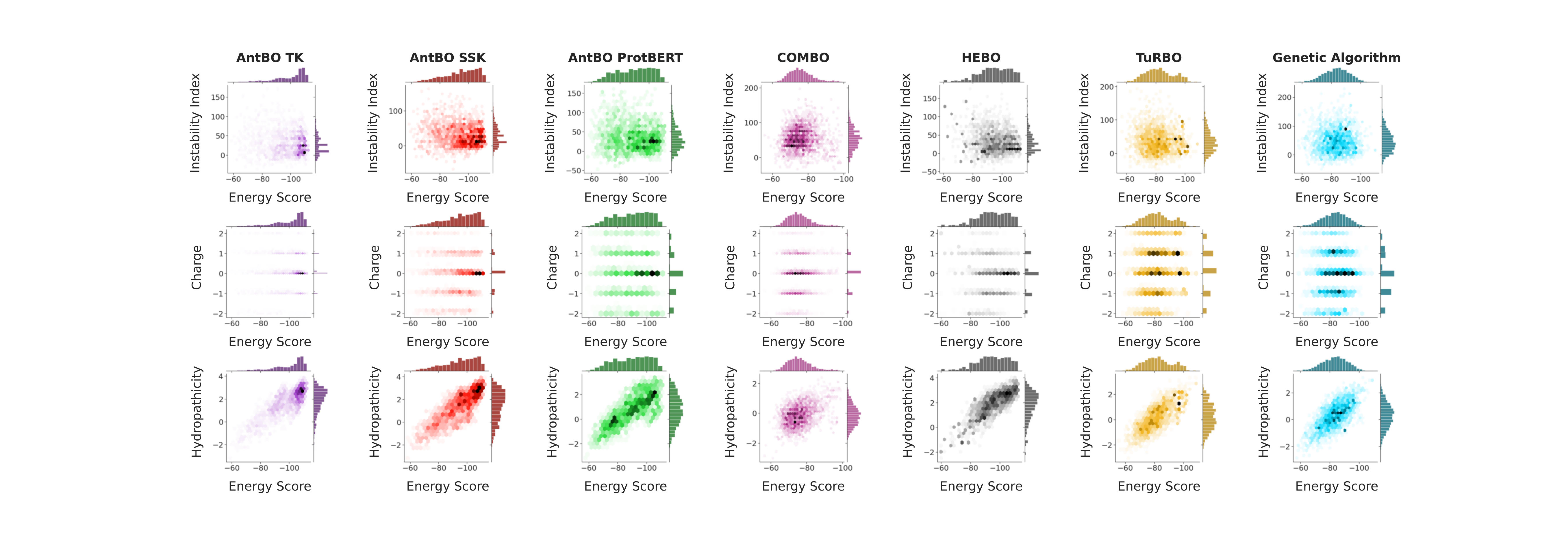}}
    \subfigure[1FBI (X)]{\includegraphics[clip, trim=11cm 3.5cm 9cm 3.5cm, width=0.9\textwidth]{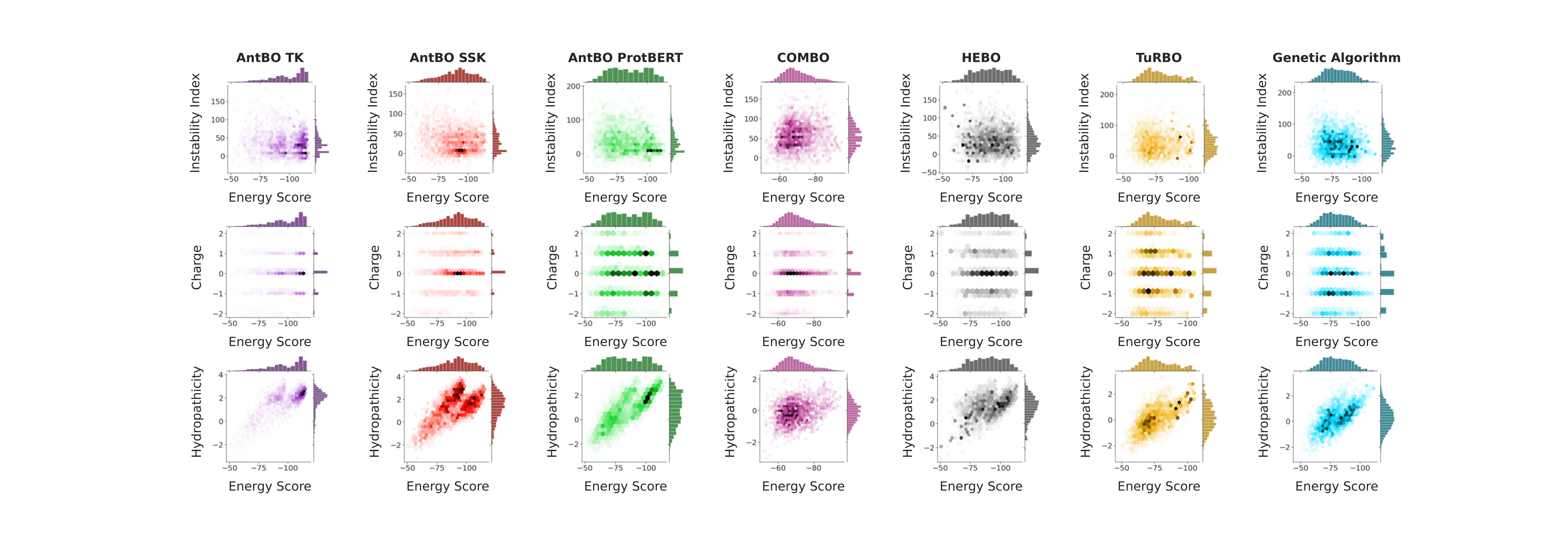}}
    \subfigure[1H0D (C)]{\includegraphics[clip, trim=11cm 3.5cm 9cm 3.5cm, width=0.9\textwidth]{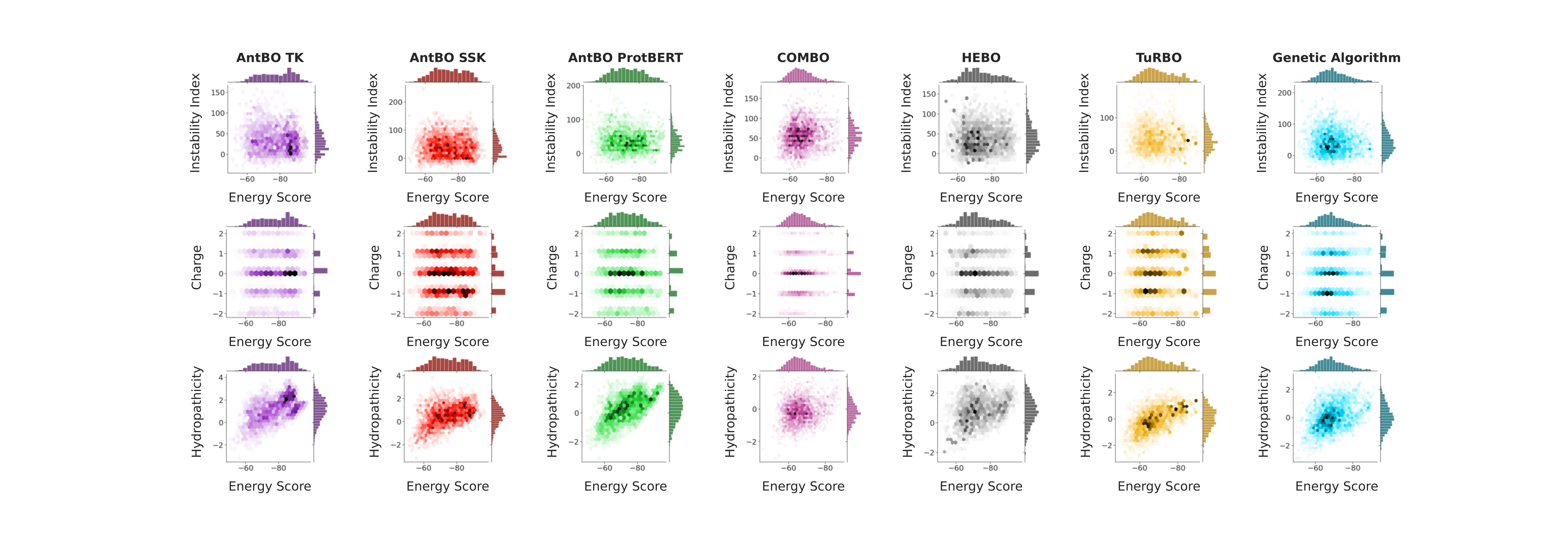}}
    \subfigure[1NSN (S)]{\includegraphics[clip, trim=11cm 3.5cm 9cm 3.5cm, width=0.9\textwidth]{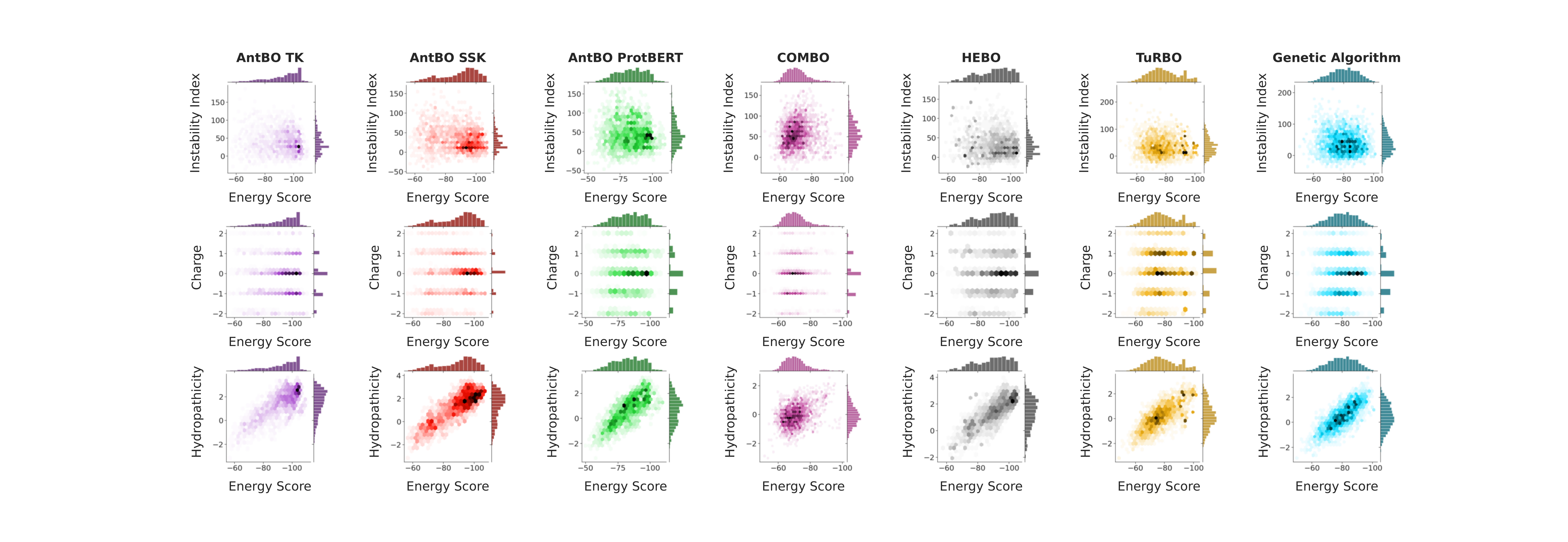}}
    \caption{Diversity of developability scores. We report the scores using the ten runs of random seeds.}
    \label{fig:dev_scores1}
\end{figure}

\begin{figure}[ht!]
    \centering
        \subfigure[1OB1 (C)]{\includegraphics[clip, trim=11cm 3.5cm 9cm 3.5cm, width=0.9\textwidth]{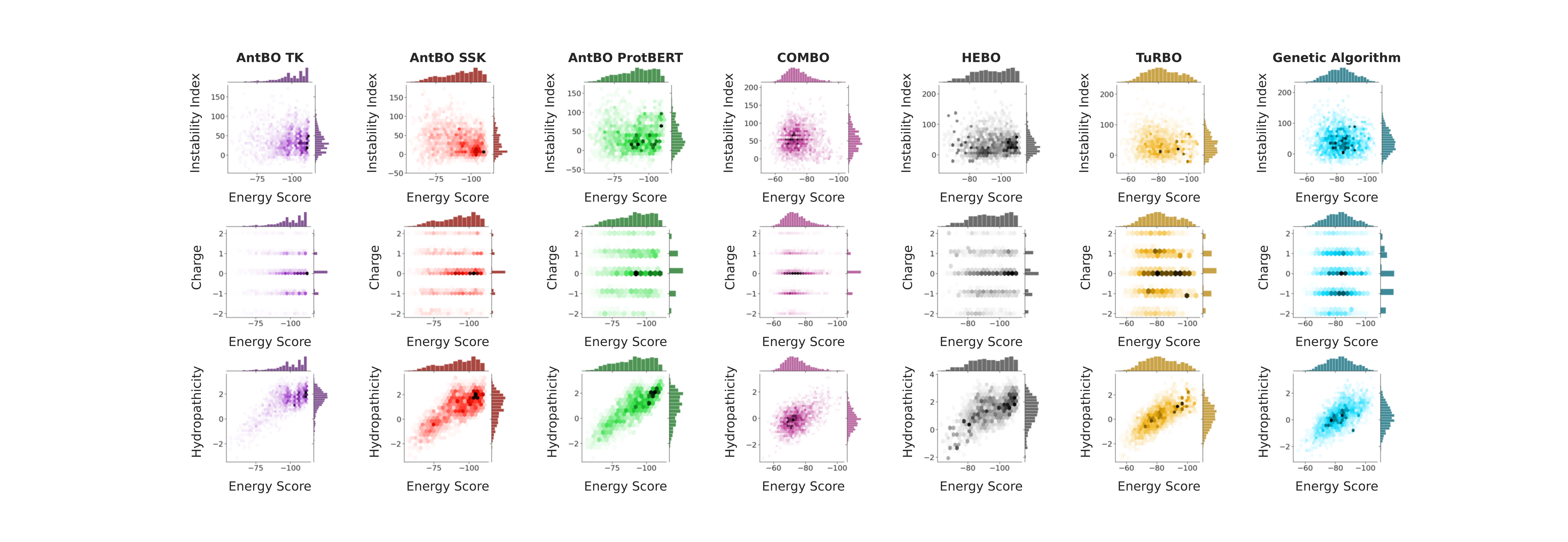}}
    \subfigure[1S78 (B)]{\includegraphics[clip, trim=11cm 3.5cm 9cm 3.5cm, width=0.9\textwidth]{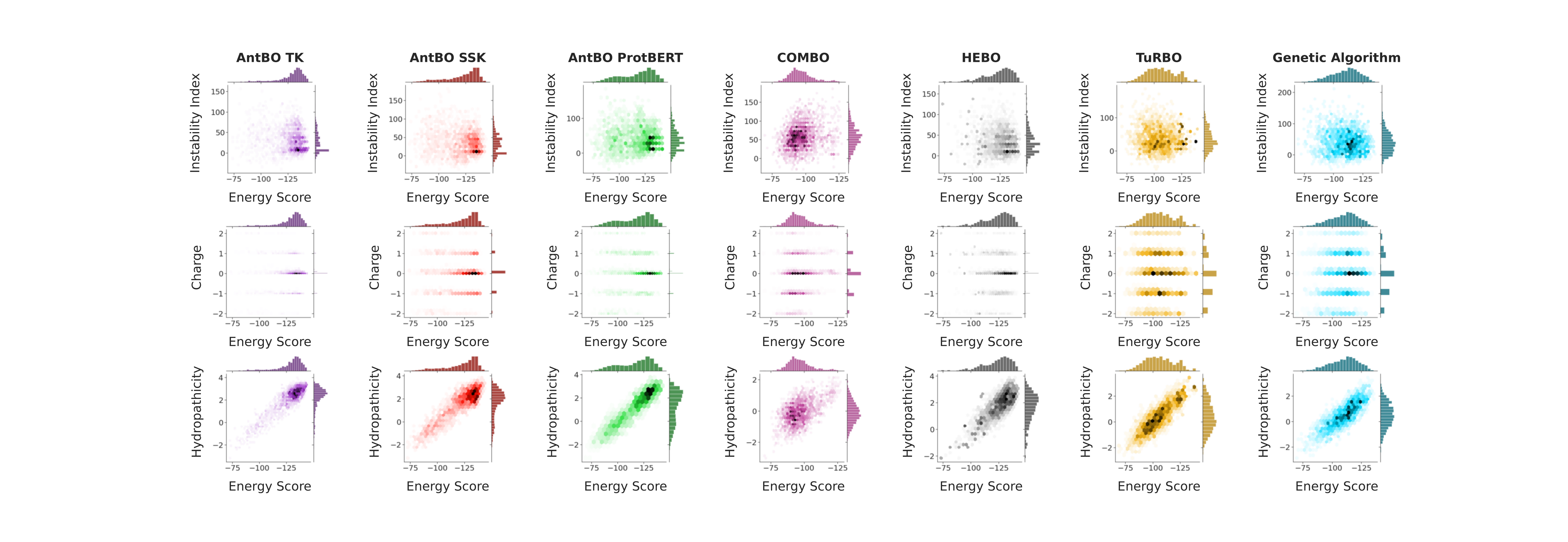}}
    \subfigure[1WEJ (F)]{\includegraphics[clip, trim=11cm 3.5cm 9cm 3.5cm, width=0.9\textwidth]{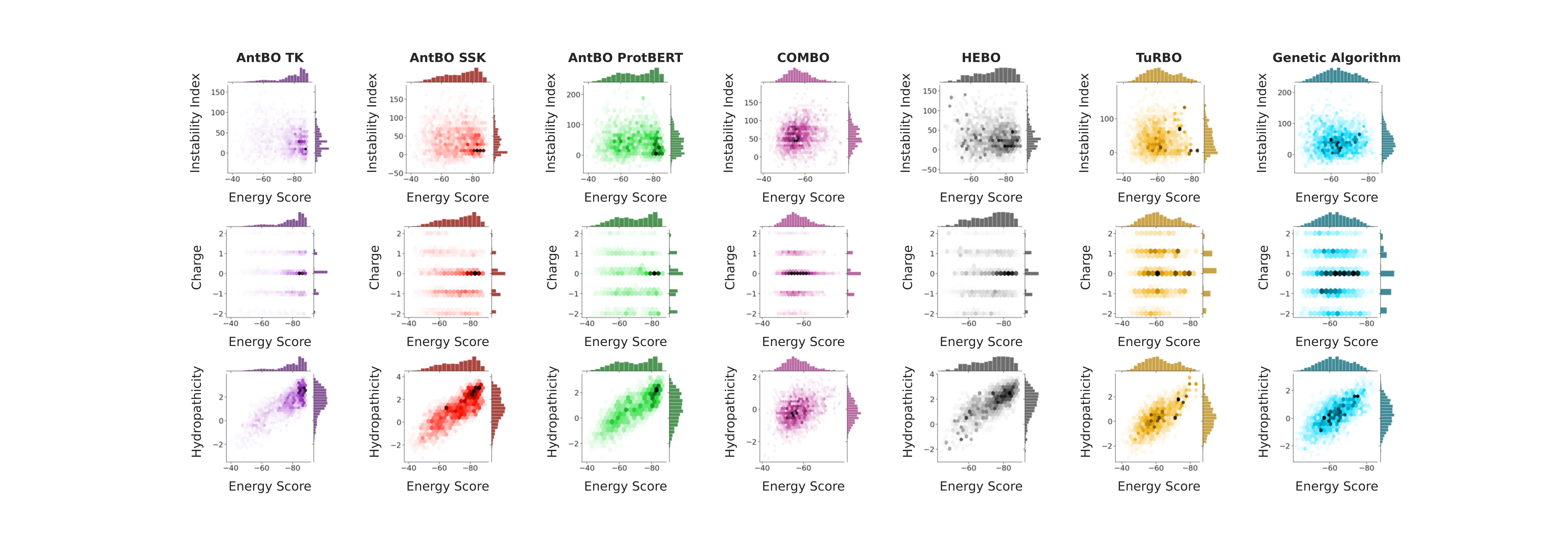}}
    \subfigure[2JEL (P)]{\includegraphics[clip, trim=11cm 3.5cm 9cm 3.5cm, width=0.9\textwidth]{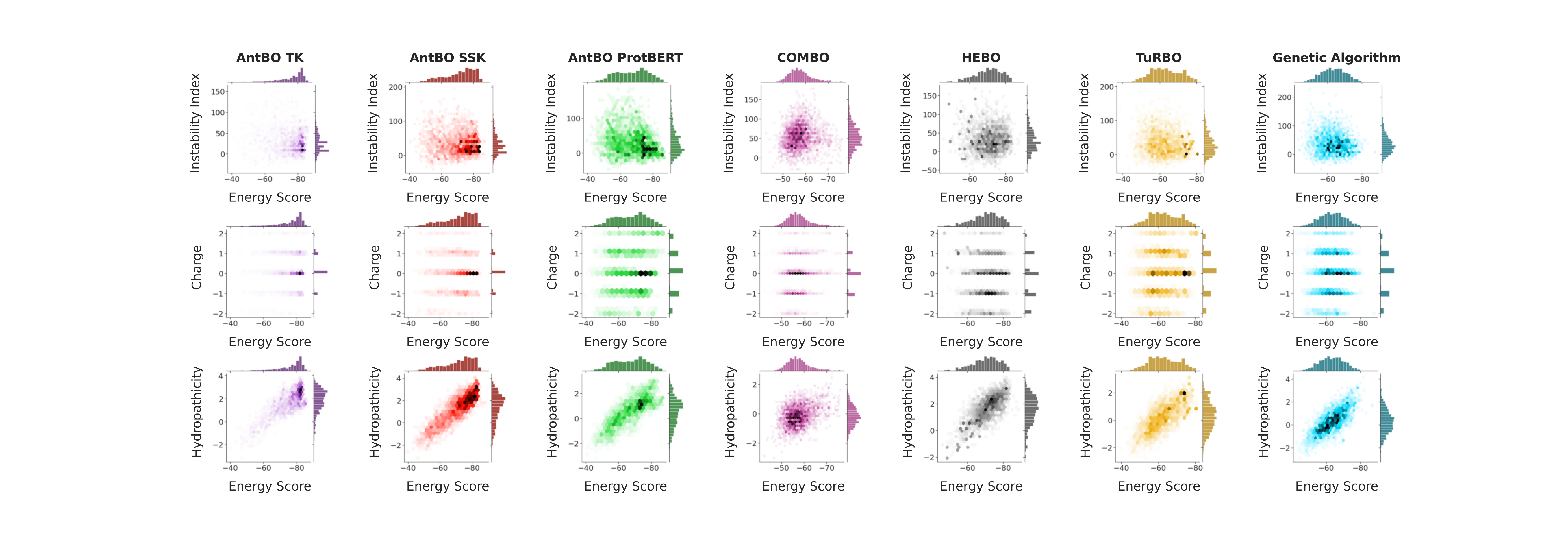}}
    \caption{Diversity of developability scores. We report the scores using the ten runs of random seeds.}
    \label{fig:dev_scores2}
\end{figure}

\begin{figure}[ht!]
    \centering
        \subfigure[2YPV (A)]{\includegraphics[clip, trim=11cm 3.5cm 9cm 3.5cm, width=0.9\textwidth]{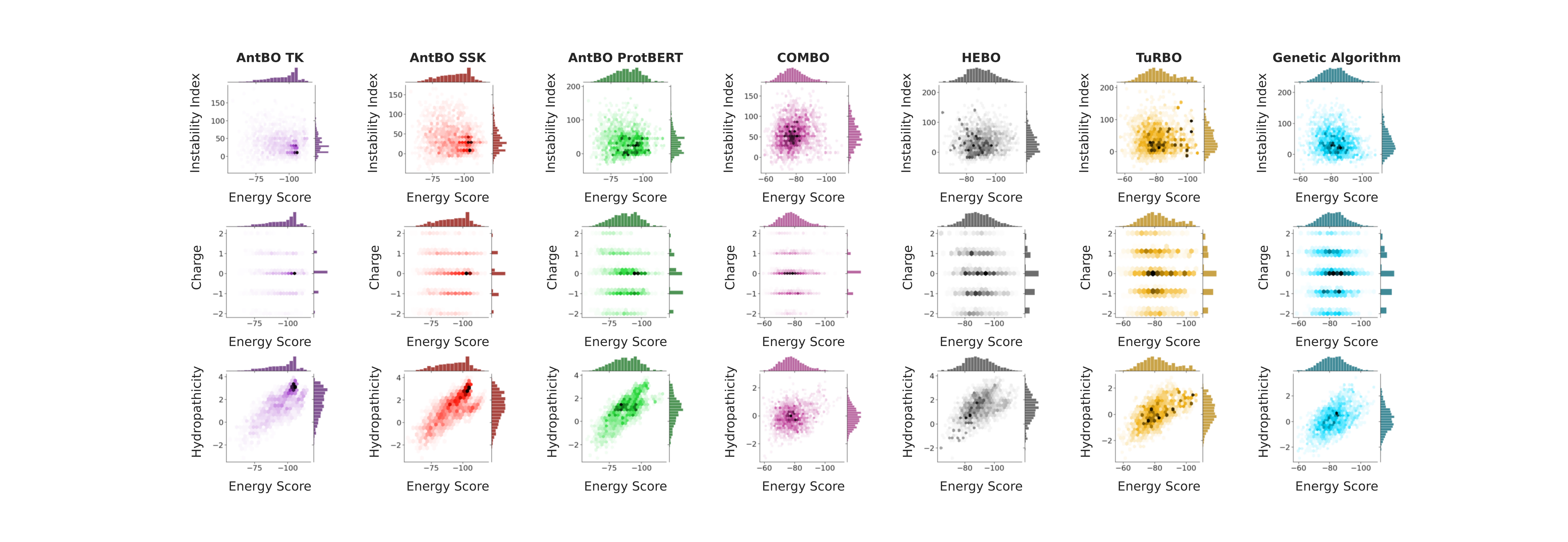}}
    \subfigure[3RAJ (A)]{\includegraphics[clip, trim=11cm 3.5cm 9cm 3.5cm, width=0.9\textwidth]{results/dev_scores/dev_scores_1S78_B.pdf}}
    \subfigure[3VRL (C)]{\includegraphics[clip, trim=11cm 3.5cm 9cm 3.5cm, width=0.9\textwidth]{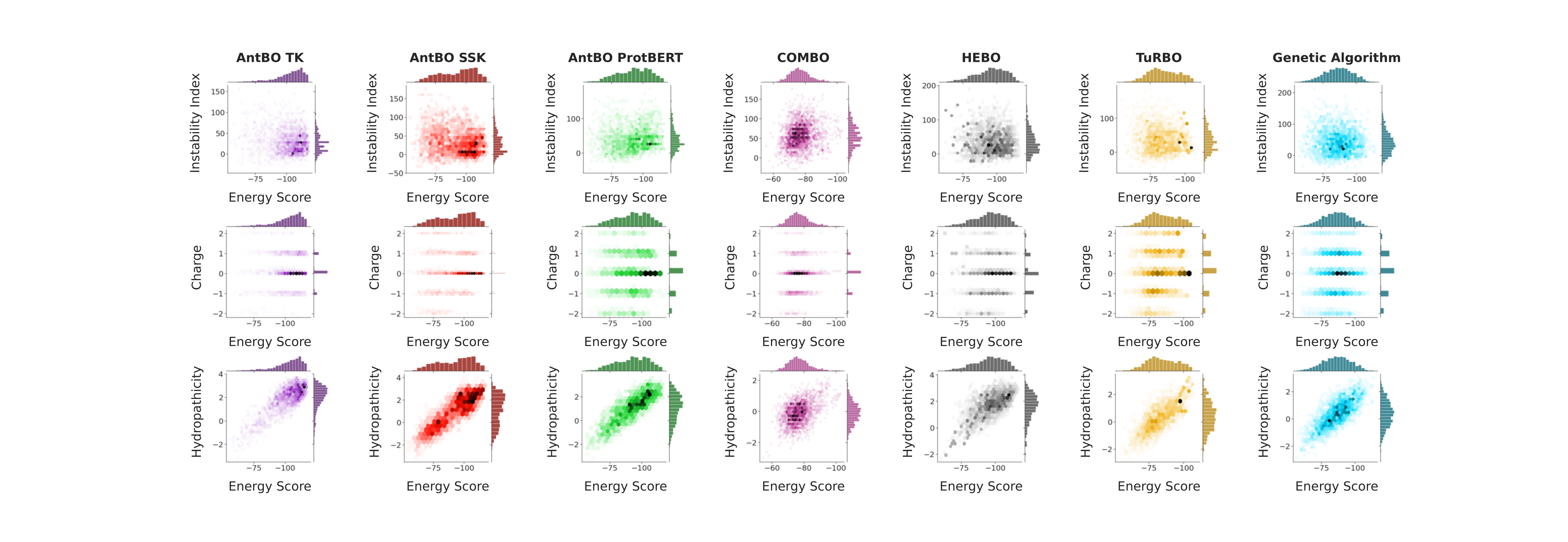}}
    \caption{Diversity of developability scores. We report the scores using the ten runs of random seeds.}
    \label{fig:dev_scores3}
\end{figure}

We next look into the range of developability scores of the rest of $11$ core antigens identified in Section~\ref{subsec:resultsmain} of the paper. Similar to analysis on SARS-COV virus we report the diversity plots for these antigens in Figure~(\ref{fig:dev_scores1},\ref{fig:dev_scores2},\ref{fig:dev_scores3}). A diverse range of scores is preferred. We observe \texttt{AntBO} is consistent in designing antibodies with stable structure.

It is known for a given antigen, several antibodies can achieve a similar binding affinity value while differing in residues in the primary structure~\cite{akbar2021silico}. We run multiple trials of \texttt{AntBO} with a distinct set of initial data points used for fitting the GP surrogate model. This nonoverlapping exploration allows local search (in the acquisition
maximisation step) to follow a different trajectory on the optimisation landscape leading to a distinct local optimum. In Table~\ref{tab:antboCDRH3s}, we report the binding energy and CDRH3 sequence obtained for twelve antigens under different trials.

\begin{figure}
    \centering
    \includegraphics[width=0.6\columnwidth, height=16cm, trim={0 2cm 8cm 0},clip]{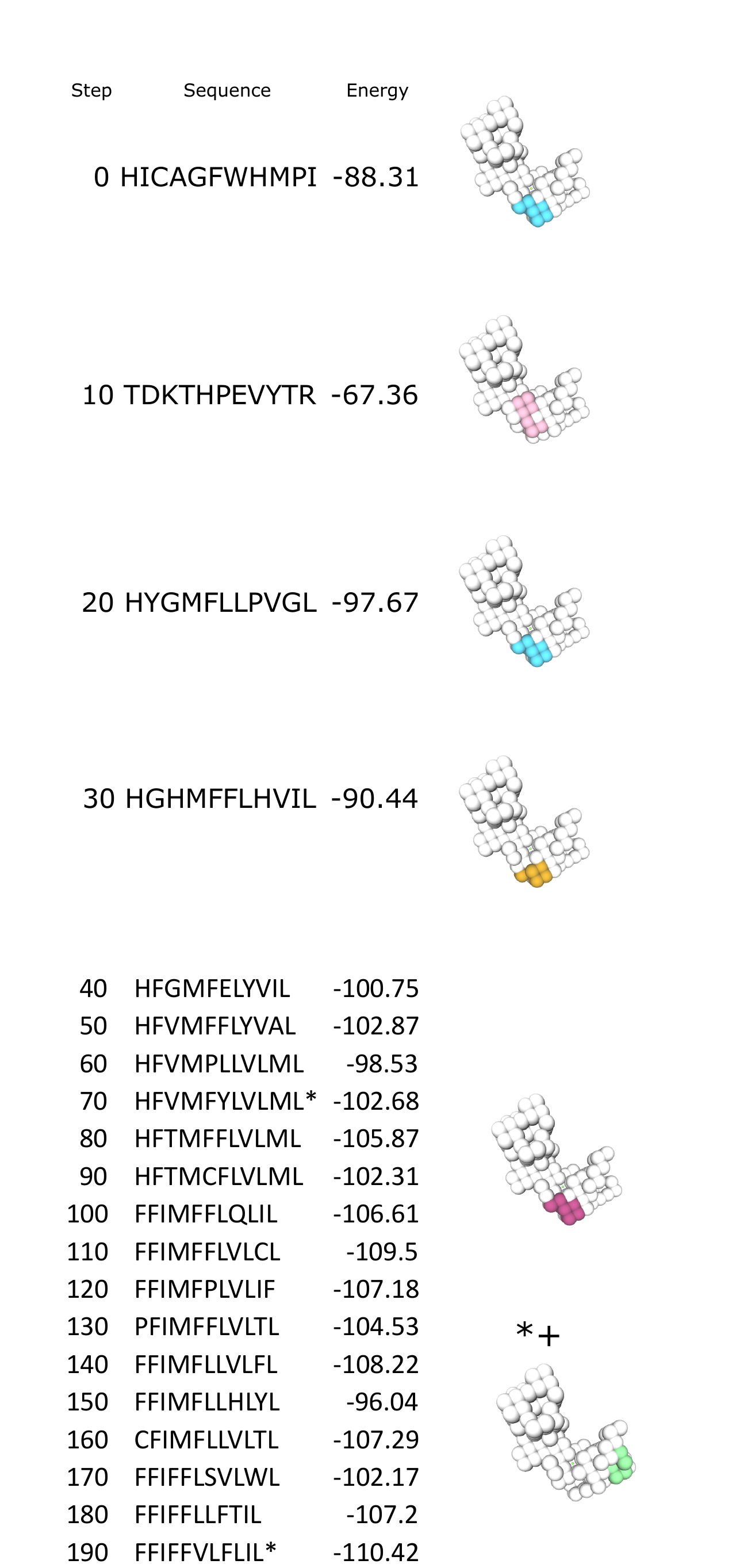}
    \caption{An example of a trajectory of sequences every ten steps generated by \texttt{AntBO}, annotated with their respective binding affinity. The structures of sequences are shown on the right. Each structure is denoted by a different colour, and from steps 40 to 197, the sequences shared the same binding structure (in  purple). Additionally, two sequences (70 and 190, marked with an asterisk) add an equally optimal binding structure (i.e., two binding modes), shown in green.}
    \label{fig:3dvisual}
\end{figure}

{\subsection{Visualisation of trajectory for antigen 1ADQ\textunderscore A}
Figure~\ref{fig:3dvisual} shows the trajectory of protein designs for an antigen 1ADQ\textunderscore A in every 10 steps. We observe that \texttt{AntBO} first explores sequences with different binding structures, later converges into regions of sequence space that contains antibodies of same binding mode and iteratively improves binding affinity by mutations that preserve the binding structure. We provide the sequence trajectories of all antigens in our codebase under the directory ``results\textunderscore data/". The instructions for the 3D visualisation of the trajectory are also provided in the codebase.}

\subsection{All methods affinity categories on core antigens}

\subsection{Binding energy vs protein design on 177 antigens}

% \begin{figure}
% \centering
% \subfigure[]{\includegraphics[width=0.9\linewidth]{results/binding_vs_funct_evals/binding_vs_funct_evals_app_6.pdf}}
% \subfigure[]{\includegraphics[width=0.9\linewidth]{results/binding_vs_funct_evals/binding_vs_funct_evals_7.pdf}}
% \caption{Binding energy vs number of protein designs.}
% \label{fig:binding_vs_funct_evals_app_7}
% \end{figure}

\subsection{Developability scores vs binding energy on core antigens}


% \end{thebibliography}
\end{document}